# Base pair interactions and hybridization isotherms of matched and mismatched oligonucleotide probes on microarrays

Hans Binder*, Stephan Preibisch, Toralf Kirsten

Interdisciplinary Centre for Bioinformatics, University of Leipzig

* corresponding author: Interdisciplinary Centre for Bioinformatics of Leipzig University, D-4107 Leipzig, Haertelstr. 16-18, binder@izbi.uni-leipzig.de, fax: ++49-341-14951-19

**Abstract**

The microarray technology enables to estimate the expression degree of thousands of genes at once by the measurement of the abundance of the respective messenger RNA. This method is based on the sequence specific binding of RNA to DNA probes and its detection using fluorescent labels. The raw intensity data are affected by the sequence-specific affinity of probe and RNA for duplex formation, by the background intensity due to non-specific hybridization at small transcript concentrations and by the saturation of the probes at high transcript concentration owing to surface adsorption. We address these issues using a binding model which describes specific and non-specific hybridization in terms of a competitive two-species *Langmuir* isotherm and DNA/RNA duplex formation in terms of sequence-specific, single-base related interactions. The GeneChip microarrays technology uses pairs of so-called perfect match (PM) and mismatch (MM) oligonucleotide probes to estimate the amount of non-specific hybridization. The mean affinity of the probes decrease according to PM(specific)>MM(specific)>>PM(non-specific)≈MM(non-specific). The stability of specific and non-specific DNA/RNA duplexes is mainly determined by Watson Crick (WC) pairings. Mismatched self complementary pairings in the middle of the MM sequence only weakly contribute to the duplex stability. The asymmetry of base pair interaction in the DNA/RNA hybrid duplexes gives rise to a duplet-like symmetry of the PM-MM intensity difference at dominating non-specific hybridization and a triplet-like symmetry at specific hybridization. The signal intensities of the PM and MM probes and their difference are assessed in terms of sensitivity and specificity. The presented results imply the refinement of existing algorithms of probe level analysis to correct microarray data for non-specific background intensities and saturation on the basis of the probe sequence.

## Introduction

Microarray chips consist of DNA oligomers with up to several hundreds of thousand different sequences that are immobilized onto a support such as glass, silicon or nylon membrane in a spot-like arrangement. They provide a powerful functional genomics technology, which permits the expression profiling of ten thousands of genes in parallel [1,2]. The working principle of this technology is based on duplex formation (hybridization) between target messenger RNA extracted from cell lines or tissues on one hand and complementary DNA nucleotide strands grafted to the chip (the reporter or probe molecules) on the other hand. Formed duplexes are detected using fluorescent or radionucleotide labels. Each spot on the chip consists of oligomers of one sequence. It is therefore representative for a certain gene and probes the abundance of the respective RNA transcript.

Different types of DNA arrays are designed for RNA profiling, which differ by the type of probes (cDNA or synthetic oligonucleotides) and by the DNA density on the array (see e.g. ref. [3]). So called high-density-oligo-nucleotide-arrays (HDONA) are produced by a photolithographic technology, which allows synthesis of oligonucleotide sequences on the chip surface in an extremely high density. This way $10^5$-$10^6$ different probe spots can be localized on one microarray of an area of about one squared centimetre [4]. The probe intensity, i.e. the integral fluorescence intensity of each probe spot, is related to the amount of bound, fluorescently labelled RNA, which in turn serves as a measure of the concentration of complementary RNA in the sample solution used for hybridization and thus of the expression degree of the respective gene.

HDONA arrays of the so called GeneChip type (Affymetrix Inc., Santa Clara) use so called probe sets of 11 (in some cases up to 20) different 25meric reporter sequences for each gene [4]. The processing of a set of several fluorescence intensities per gene is expected to improve the reliability of the method. Note that the sample RNA is cleaved into fragments with a length of several dozen nucleotides before hybridization. The RNA fragments referring to different regions of the target gene are expected to bind virtually independently to the oligonucleotide probes of one set.

The target RNA for each probe constitutes only a fraction of the total RNA in the sample solution used for hybridization. A considerable amount of RNA involves other sequences than the intended target. Unfortunately these non-specific transcripts compete with the target RNA for duplex formation with the probes. This way they also contribute to the signal intensities due to non-specific binding. The lack of specificity raises a serious problem for the analysis of microarray data because the residual "chemical background" intensity is not related to the expression degree of the gene of interest and therefore distorts the signal of specifically bound target RNA.

To deal with this problem each probe sequence on GeneChip micoarrays is present in two modifications called perfect match (PM) and mismatch (MM) probes. The sequence of the PM is taken from the gene of interest and thus it is complementary to a 25mer in the RNA target sequence. The sequence of the MM is identical with that of the PM probe except the position in the middle of the sequence where the "middle base" is replaced by its complement. The MM are designed as reporters for non-specific hybridization that measure the intensity of the "chemical background", i.e. of bound RNA not referring to the target gene. The MM signal provides a potential correction of the respective PM intensity for the contribution of non-specific binding.

The idea behind the paiwise design of probes is based on the assumption that non-specific transcripts bind with virtually identical affinity to the PM and MM probes of one pair whereas the target RNA is expected to hybridize the MM with considerable less affinity due to the mismatched base pairing in accordance with "conventional" hybridization thermodynamics of DNA/RNA duplexes in solution [5]. The lower stability of mismatched probe/target duplexes predicts a systematically equal or smaller intensity of the MM probes. It was however found, that a considerable fraction of the MM probes fluoresces with higher intensity than the paired PM [6]. Consequently subtracting MM from PM intensities as a way of correcting the PM intensities for non-specific binding seems not always appropriate [7,8]. As a consequence the "mysterious" MM were either completely ignored in signal analysis algorithms [9,10] or they are considered in an empirical fashion to exclude "bad" probes from the analysis [4].

Hence, one important question for GeneChip data analysis is how to include the MM intensities adequately? This more technical issue, in turn, requires the detailed understanding of the basic rules of oligonucleotide duplex formation on microarrays and, in particular, of the hybridization mechanism of matched and mismatches microarray probes with specific and non-specific RNA transcripts on the level of base pairings. The effect of competitive hybridization of specific and non-specific RNA fragments on the thermodynamically attainable performance of DNA chips can be quantified in terms of the hybridization (or binding) isotherms of the PM and MM probe spots. The isotherms provide a basic characteristic of the probes because they relate the degree of the hybridization to the bulk RNA composition, and thus to the expression degree. The instrumental response characteristic must in addition consider the effect of selective labelling which produces the fluorescence intensity measured by the detector.

The present paper addresses these issues in terms of a hybridization model, which explicitly considers the RNA concentration and the amount of specific and non-specific transcripts in the sample solution on the one hand and the sequence of the oligonucleotide probes and especially their middle base on the other hand. The theoretical results are compared with microarray intensity data, which were taken from a calibration experiment provided by Affymetrix.

## Microscopic model of hybridization on microarrays

### Binding affinity and the intensity of oligonucleotide probes

Gene expression analysis by means of high-density-oligonucleotide-array (HDONA) chips is based on the sequence specific binding of RNA fragments to oligonucleotide probes and its measurement using fluorescenct labels. Affymetrix uses short 25mers as perfect match (PM) probes the sequence, $\xi^{PM}$, of which is complementary to a fragment of the consensus sequence of the respective target gene, $\xi^{T}$ [4]. The probe and target sequences are given by strings of $N_b = 25$ letters (A, T, G or C), e.g., $\xi^P$= 3'-ACCCAG…T-5' and $\xi^T$= 3'-u*gggu*c*…a-5' (uppercase letters refer to the DNA probe, lower case letter refer to the RNA, the asterisk denotes labelling).

The PM probe intends to bind the target RNA via the Watson Crick (WC) pairings A-u*, T-a, G-c* and C-g. The respective association constant for duplex formation, $K_p^{PM,S} = K_p^b(\xi_p^{PM}\xi_p^T)$ (the index p denotes the probe number), quantifies the strength of specific binding between target and probe according to the binding reaction $(\xi^{PM}\xi^T) \leftrightarrow \xi^{PM} + \xi^T$. The association constant for duplex formation of the mismatch (MM) probe with the target, $K_p^{MM,S} = K_p^b(\xi_p^{MM}\xi_p^T)$, characterizes the affinity of target RNA for specific binding despite the fact that the middle base of the MM probe disables WC pairings. Instead, the 13$^{th}$ base is assumed to form the respective self complementary (SC) pair with the target RNA, $\underline{A}$-a, $\underline{T}$-u*, $\underline{G}$-g or $\underline{C}$-c*.

The sample solution used for hybridization usually contains a large number of RNA fragments with sequences differing from that of the target, i.e. $\xi \neq \xi^T$. Also these non-specific RNA fragments bind in significant amounts to the probes despite the fact that probe and DNA only partly match each other via WC pairings. The respective association constants, $K_p^P(\xi_p^P\xi)_{\xi \neq \xi^T}$ quantify the affinity of the probe (P=PM, MM) for duplex formation with non-specific RNA fragments of sequence $\xi$ according to the reaction $(\xi^P\xi) \leftrightarrow \xi^P + \xi$. The mean binding affinity of the probe for non-specific hybridization is given by the concentration-weighted average over the binding constants of this "cocktail" of RNA sequences, $K_p^{P,NS} = \langle K_p^P(\xi_p^P\xi) \rangle_{\xi \neq \xi^T} \equiv \sum_{\xi \neq \xi^T} c_{RNA}(\xi) \cdot K_p^P(\xi_p^P\xi) / \sum_{\xi \neq \xi^T} c_{RNA}(\xi)$. The non-specific fragments are expected to bind with lower affinity to the probe compared with the target RNA owing to the smaller number of WC pairings. The ratio $r_p^P = K_p^{P,NS} / K_p^{P,S} < 1$ specifies the mean relative binding strength of the probe for non-specific hybridization compared with that of specific binding with the target sequence, $\xi^T$ [11].

The amount of probe-bound RNA is detected by means of fluorescent labels, which are linked to the uracyls (u*) and cytosines (c*). The respective fluorescence intensity per probe spot measured by the detector can be described by [11]

$$I_p^P \approx F_{chip} \cdot N_p^{F,S} \cdot K_p^{P,S} \cdot \left[ c_{RNA}^S + c_{RNA}^{NS} \cdot r_p^P \cdot r_p^{F,P} \right] \cdot S_p^P \qquad (1)$$

if one neglects the optical background. Essentially four factors affect the signal intensity of a particular probe according to Eq. 1:

(i) The binding "strength" (or affinity) of the DNA probe for duplex formation with the RNA fragments upon hybridization determines the amount of RNA that binds to the probe. It is characterized by the binding constant of specific hybridization, $K_p^{P,S}$ and the mean relative strength of non-specific binding, $r_p^P$. In Eq. 1 the binding equilibria between the probe and all relevant non-specific RNA sequences are replaced by one equilibrium between the probe and a characteristic non-specific transcript, which is characterized by the mean binding constant $K_p^{P,NS}$. In other words, the cocktail of non-specific RNA fragments is assumed to act like a single species in accordance with previous treatments of cross hybridization [12]. Equation 1 further considers saturation of the probes with specific and non-specific RNA fragments, which both compete for the free binding sites provided by the monomeric oligomers according to a competitive two-species Langmuir isotherm, $S_p^P = \left(1 + K_p^{P,S} \cdot \left[c_{RNA}^S + c_{RNA}^{NS} \cdot r_p^P\right]\right)^{-1}$ (see also Eq. 1).

(ii) The fluorescence "strength" (or yield) of the hybridized RNA determines the emitted intensity per bound transcript. It is roughly related to the amount of labelling, which is given by the mean number of fluorescently labelled cytosines and uracyls in the sequence of the respective fragment of bound RNA, $N_p^{F,S} = N_p^{c*} + N_p^{u*}$. The ratio $r_p^{F,P} = N_p^{F,P,NS}/N_p^{F,S}$ specifies the relation between the amount of labelling of non-specifically and specifically hybridized probes. Note that the specific target RNA fragment is identical for the PM and MM probes of one probe pair, whereas the non-specific RNA effectively differs by one base (see below).

(iii) The total concentration of RNA fragments in the sample solution used for hybridization, $c_{RNA}^{tot} = \sum_\xi c_{RNA}(\xi)$, is directly related to the amount of binding according to the mass action law. It splits into the concentration of target RNA (specific transcripts), $c_{RNA}^S = c_{RNA}(\xi^T)$, and into the concentration of non-specific transcripts involving other sequences than the intended target, $c_{RNA}^{NS} = c_{RNA}(\xi \neq \xi^T)$.

(iv) The chip specific constant $F_{chip}$ considers the detection "strength" of the technique. It considers aspects of chip fabrication such as the number and density of oligonucleotides per probe spot, the sensitivity of the imaging system and factors due to the performance of the experiment, e.g. the yield of labelling.

Note that the microarray experiment intends to measure the expression degree of the target gene in terms of $c_{RNA}^S$, the concentration of specific transcript. Signal analysis consequently requires the correction of the measured intensity for the effect of labelling, the chip specific constant, and most importantly, for saturation and non-specific hybridization.

## Mean binding isotherm and the free energy of duplex formation

Let us split the log-intensity of each probe into a mean value, $<\log I_p^P>_\Sigma$, averaged over an appropriate ensemble of probes $\Sigma$ referring to one concentration of specific transcripts (i.e. $c_{RNA}^S$=const) on one hand, and an incremental contribution, which reflects the individual properties of the selected probe, $Y_p^P \equiv \Delta \log I_p^P$, on the other hand, i.e.,

$$\log I_p^P = <\log I_p^P>_\Sigma + Y_p^P \qquad (2)$$

In this work we use two options for ensemble averaging. Firstly, in the so called spiked-in experiment (see below) RNA transcripts of selected probes were titrated onto a series of chips in well-defined concentrations, $c_{RNA}^S = c_{spiked-in}$. In this case the respective ensemble of probe intensities referring to explicitly known concentration, $c_{spiked-in}$=const ($\Sigma$=spiked-in), were taken from different chips. Alternatively, one can pool all probes of a probe set ($\Sigma$=set) together because they refer to one gene and consequently to one target concentration with $c_{RNA}^S$=const., which is however apriori unknown. In this case averaging was performed over probes from one chip.

The mean intensity can be described by an effective binding isotherm adapted from Eq. 1,

$$\left\langle \log I^P \right\rangle_\Sigma \approx \log F + \log K_0^{P,S} + \log\left[ c_{RNA}^S + c_{RNA}^{NS} \cdot r_0^P \right] + \log S_0^P$$

with

$$\log F = \log F_{chip} + \log N_0^F \quad , \quad S_0^P = \left(1 + K_0^{P,S} \cdot \left[c_{RNA}^S + c_{RNA}^{NS} \cdot r_0^P\right]\right)^{-1} \quad and \quad r_0^P = K_0^{P,NS} / K_0^{P,S} \qquad (3)$$

The effective constants $\log K_0^{P,h} = <\log(K_p^{P,h})>_\Sigma$ (h=S, NS) and $\log N_0^F = <\log(N_p^{F,S})>_\Sigma$ represent mean values over all considered probes. Equation 3 assumes that the log-intensity average is a function of these effective values and that hybridized PM and MM probes are equally labelled on the average.

The binding constant of the probes provides the respective Gibbs free energy of duplex formation, $\Delta G_p^{P,S} = \mu^0_{duplex}(\xi_p^P \xi_p^T) - ( \mu^0_{DNA}(\xi_p^P) + \mu^0_{RNA}(\xi_p^T))$ and $\Delta G_p^{P,NS} = <\mu^0_{duplex}(\xi_p^P \xi) - ( \mu^0_{DNA}(\xi_p^P) + \mu^0_{RNA}(\xi))>_\xi$, where the $\mu^0$ denote the respective standard chemical potentials of the reactants and of the duplex. The free energy of duplex formation can be decomposed into a sum of base and positional dependent contributions,

$$\Delta G_p^{P,h} = -RT \ln\left(W \cdot K_p^{P,h}\right) = RT \ln 10 \cdot \sum_{k=1}^{N_b} \varepsilon_k^{P,h}(\xi_{p,k}^P) \quad with \quad h = S, NS \qquad (4)$$

where $\xi_{p,k}^P$ denotes the nucleotide base at position k of the probe sequence, R and T are the gas constant and the temperature, respectively, and W is the cratic contribution accounting for the mixing entropy [13].

The free energy terms can be further split into a base independent mean value averaged over the chosen ensemble of probes and into a base dependent contribution in analogy with Eq. 2

$$\varepsilon_k^{P,h}(B) = \varepsilon_{0,k}^{P,h} + \Delta \varepsilon_k^{P,h}(B) \quad with \quad B = A, T, G, C$$

$$and \quad \sum_{k=1}^{N_b} \varepsilon_{0,k}^{P,h} = -\log K_0^{P,h} + const \qquad (5)$$

In general, the hybridization at the surface of a DNA chip differs from the Langmuir scenario in that both the adsorbates (the targets) and the surface (the probe layer) are charged. As a result the free energy of duplex formation incorporates electrostatic terms, which depend on the amount of bound RNA [12,14]. For this situation the binding constant has to be supplemented by a concentration dependent exponential factor, which considers the progressive depletion of the free adsorbate near the surface owing to electrostatic repulsion between bound and free species. This effect gives rise to a saturation-like behaviour where further binding with increasing bulk concentration of the adsorbate is effectively hampered by always bound species. Despite these limitations we will use the Langmuir form as a good approximation because it provides a satisfactory description of the used experimental data (see below and also [15] [16,17]). The resulting binding constants (and free energies) must be interpreted as apparent values that include the electrostatic contribution.

The competitive two-species Langmuir isotherm assumes two discrete energetic states for specific and non-specific hybridization (see above). The explicit consideration of a continuous distribution of binding free energies due to the heterogeneity of RNA sequences can be achieved by the replacement $c \cdot K \rightarrow (c \cdot K)^a$ (with the exponent a < 1) in the respective Langmuir-type isotherm [18]. Note however that even the most critical application of the used Langmuir form to the average over all probes (Eq. 3) actually provides a good description of the experimental data (see below). We therefore judge this simpler Langmuir-version as the adequate approach in this work.

**The sensitivity of the oligonucleotide probes**

The incremental contribution to the intensity

$$Y_p^P = \log I_p^P - \left\langle \log I_p^P \right\rangle_\Sigma \quad , \quad P = PM, MM \tag{6}$$

defines the sensitivity of the respective probe, which, in a first order approximation, characterizes its ability to detect a certain amount of RNA independently of the experimental conditions given by the chip specific factor $F_{chip}$. Note that the transformation according to Eq. 6 cancels out all factors to the intensity, which are common for the chosen ensemble of probes. Our definition of the sensitivity for the special case of oligonucleotide probes on GeneChip microarrays is adapted from the general definition of the IUPAC for analytical techniques, which identifies the sensitivity with the measured response per concentration increment (see [19] and references cited therein).

Insertion of Eq. 6 into Eq. 1 shows that the probe sensitivity additively decomposes into terms due to the binding affinity and fluorescence [11],

$$Y_p^P \approx Y_p^{P,b} + Y_p^{P,F}$$
$$Y_p^{P,b} = \log\left(K_p^{P,S} \cdot \left[c_{RNA}^S + c_{RNA}^{NS} \cdot r_p^P \cdot r_p^{P,F}\right] \cdot S_p^P\right) - \left\langle \log\left(K_p^{P,S} \cdot \left[c_{RNA}^S + c_{RNA}^{NS} \cdot r_p^P \cdot r_p^{P,F}\right] \cdot S_p^P\right) \right\rangle_\Sigma \tag{7}$$
$$Y_p^{P,F} = \log\left(N_p^{F,S}\right) - \left\langle \log\left(N_p^{F,S}\right) \right\rangle_\Sigma$$

**Positional dependent single base (SB) model of the sensitivity**

Positional dependent SB models were recently used to predict microarray probe intensities [20,21]. In our notation the SB model decomposes the sensitivity of each probe into a sum of sensitivity contributions $\sigma_k^P(\xi_k^P)$, depending on the base at position $k = 1\ldots N_b$ of the probe sequence, $\xi_k^P$,

$$Y_p^{P,SB} = \sum_{k=1}^{N_b} \sum_{B=A,T,G,C} \sigma_k^P(B) \cdot \left(\delta(B, \xi_{p,k}^P) - f_k^\Sigma(B)\right) \tag{8}$$

Here $\delta$ denotes the Kronecker delta ($\delta(x,y)=1$ if $x=y$ and $\delta(x,y)=0$ if $x \neq y$). The term $f_k^\Sigma(B)$ is the fraction of base B at position k in the considered ensemble of probes, or, in other words, the probability of occurrence of letter B at position k in the ensemble. The SB sensitivity contributions at a given position spread symmetrically about zero, i.e., they are restricted to the condition

$$\sum_{B=A,T,G,C} \sigma_k^P(B) = 0 \quad \text{for all positions} \quad k=1\ldots N_b \quad . \tag{9}$$

The position-averaged SB sensitivity ,

$$\sigma^P(B) = \left\langle \sigma_k^P(B) \right\rangle_k \equiv \frac{1}{N_b} \sum_{k=1}^{N_b} \sigma_k^P(B) \quad , \tag{10}$$

characterizes the mean contribution of base B to the sensitivity independently of its position along the sequence. The mean over all bases in terms of absolute values,

$$\sigma^P = \left\langle |\sigma_k^P(B)| \right\rangle_{k,B} = \frac{1}{4 \cdot N_b} \sum_{k=1}^{N_b} \sum_{B=A,T,G,C} |\sigma_k^P(B)| \quad , \tag{11}$$

can be interpreted as a measure of the variability of the sensitivity of the probes due to sequence specific effects (see also Eq. 6).

The intensity of a selected probe represents the superposition of the respective ensemble averaged intensity and of the sequence specific contribution given by the SB sensitivity model (see Eqs. 6 and 8),

$$\log I_p^P \approx \left\langle \log I^P \right\rangle_\Sigma + \sum_{k=1}^{N_b} \sigma_k^P(\xi_{p,k}^P) \quad . \tag{12}$$

In the general case both, the mean intensity and the SB contributions are functions of the RNA target concentration. Let us neglect saturation for sake of simplicity. Then insertion of Eq. 3 into 12 provides

$$\log I_p^P \approx \log F + \log K_0^{P,h} + \log c_{RNA}^h + \sum_{k=1}^{N_b} \sigma_k^{P,h}(\xi_{p,k}^P) \tag{13}$$

in the limiting case of specific (h=S for $c_{RNA}^S \gg c_{RNA}^{NS} \cdot r_0^P$) and of non-specific (h=NS for $c_{RNA}^S \ll c_{RNA}^{NS} \cdot r_0^P$) hybridization, respectively. Equation 13 shows that the fit of the SB model to the sensitivities of an appropriately chosen ensemble of probes provides estimates of SB sensitivity parameters, which characterize specific and non-specific DNA/RNA probe/target duplexes.

**Fluorescence contribution**

The sensitivity of each probe divides into two additive contributions according to Eq. 7 due to (i) the binding "strength" of the RNA for duplex formation with the probe and (ii) the fluorescence "strength" of bound RNA. A relatively high binding strength consequently represents a necessary but not sufficient condition of highly sensitive probes. In addition the bound RNA must emit light with sufficiently high intensity, which in turn depends on the amount of labelling of the probe/RNA duplex. Both, the binding affinity and the fluorescence yield are functions of the base composition of the probe. It appears therefore appropriate to split the SB sensitivity into a Gibbs free energy and a fluorescence contribution,

$$\sigma_k^{P,h}(\xi_{p,k}^P) = -\Delta\varepsilon_k^{P,h}(\xi_{p,k}^P) + \Delta\varphi_k^{P,h}(\xi_{p,k}^P) \quad with \quad h = S, NS \tag{14}$$

The former term, $\Delta\varepsilon_k^{P,h}(\xi_{p,k}^P)$, characterizes the relative binding strength of the base at position k of the probe sequence $\xi_p^P$ (see also Eq. 5). The fluorescence contribution, $\Delta\varphi_k^{P,h}(\xi_{p,k}^P)$, considers the fact that not every base fluoresces owing to the specificity of labelling for cytosines and uracyls. The free energy and fluorescence contributions are assumed to meet the symmetry condition (Eq. 9).

The fluorescence intensity of a RNA target fragment is related to the number of labelled bases, c* and u*. It, in turn, depends on the number of complementary bases, B=G and A, in the PM probe sequence if one assumes binding via WC pairs. In the Supplementary Material (S1) we show that the positional dependent SB sensitivity contributions of labelled bases are enhanced whereas the contributions of non-labelled bases are decreased by a constant, positional independent increment $\Delta^F$ (see Eq. A3 in the Supplementary Material),

$$\Delta\varphi_k^{PM,S}(B) = \Delta\varphi_k^{WC}(B) \approx \begin{cases} +\Delta^F & for \quad B = A, G \\ -\Delta^F & for \quad B = T, C \end{cases} \quad with \quad \Delta^F \approx \left(1 + \sqrt{\frac{N_b^{RNA}}{N_b} - 1}\right) \cdot \frac{2}{\ln 10 \cdot N_b^{RNA}}. \tag{15}$$

This result assumes direct proportionality between the emitted fluorescence intensity and the number of potentially labelled bases in the target sequence, $I^P \propto N_p^F$. The fluorescence effectively increases the single base sensitivity of labelled A-u* and G-c* WC base pairs and decreases the sensitivity of non-labelled T-a and C-g pairs in a symmetrical fashion. For self complementary pairs (A-a, T-u*, G-g and C-c*) the relation reverses, i.e., $\Delta\varphi_k^{WC}(B) = -\Delta\varphi_k^{SC}(B)$.

The increment $\Delta^F$ depends on the total sequence length of the RNA fragments, $N_b^{RNA}$, and on the length of the probe oligomers, $N_b=25$, which is explicitly considered in the SB model. One obtains $\Delta^F \approx 0.04$ if the target length exactly matches the probe ($N_b^{RNA}=25$). The fluorescence term remains nearly constant for longer target sequences up to $N_b^{RNA} = 50$ and then it progressively decreases to $\Delta^F \approx 0.03$ for $N_b^{RNA} = 65$ and to values less than 0.02 for $N_b^{RNA} > 100$ nucleotides. Hence, a value of $\Delta^F \approx 0.04$ can be judged as an upper limit of the fluorescence contribution to the positional dependent sensitivity.

The comparison between the binding data of labelled and non-labelled oligonucleotides shows that labelling (i.e., the covalent linkage of biotinyl residues with attached fluorescent labels to the nucleotide bases) slightly but significantly decreases the binding strength of a nucleotide base by a reduced free energy increment of less than 0.05 (see Eq. 4, [22]). Hence, the fluorescence strength and the change of the free energy contribution owing to labelling obviously compensate each other at least partially with respect to their effect on the SB sensitivity.

## Data processing and parameter estimation

### Chip data

Microarray intensity data are taken from the Affymetrix' human genome Latin Square (HG U133-LS) data set available at http://www.affymetrix.com/support/technical/ sample_data/datasets.affx. These data are obtained in a calibration experiment, in which specific RNA transcripts referring to 42 genes (and thus to $N_{data}$=11x42=462 PM/MM probe pairs) were titrated in definite concentrations onto microarrays of the Affymetrix HG U133 type to study the relation between the probe intensity and the respective ("spiked-in") RNA concentration. Fourteen different concentrations ranging from 0 pM (i.e. no specific transcripts) to 512 pM were used for each probe. The experiment further uses 14 different arrays for all cyclic permutations of the spiked-in concentrations and spiked-in genes (the so-called Latin Square design). Non-specific hybridization was taken into account by adding a complex human RNA background extracted from a HeLa cell line not containing the spiked-in transcripts to all hybridization solutions. The PM and MM probe intensities were corrected for the optical background before further analysis using the algorithm provided by MAS 5.0 [4].

### Least square fits

The sensitivity coefficients of the SB model, $\sigma_k^P(B)$, were determined by means of multiple linear regression which minimizes the sum of weighted squared residuals between measured and calculated sensitivities [23], $SSQR = \sum_{p=1}^{Ndata} \omega_p^{-2} \left(Y_p^P - Y_p^{P,SB}\right)^2$. The sum runs over all considered probes $N_{data}$. The resulting system of linear equations was solved by means of single value decomposition (SVD, [24]), which guarantees the solution that meets the symmetry condition (Eq. 9).

The weighting factor, $\omega_p^2$, was estimated using the error model described in the Supplementary Material (S2), $\omega_p^2 = var(log(I_p)) = a + b/(I_p^P) + c/(I_p^P)^2$. It accounts for the increase of signal error at small intensities in a logarithmic scale. The constants a, b and c consider the noise level of the binding equilibrium, of a probe-specific stochastic term and of the optical background, respectively. They were estimated using a set of more than 3000 oligonucleotide probes present as replicates on each HG U133 chip.

# Results

**Binding isotherms and signal intensities of individual probes**

The spiked-in LS data set provides PM and MM intensities of 42 selected probe sets as a function of the concentration of specific target RNA in a constant background of non-specific hybridization. The concentration dependence of the intensity of six selected probe pairs is shown in Fig. 1. The courses are well described by Eq. 1 (compare lines and symbols, note the logarithmic scale). Accordingly, each curve is characterized by two model parameters, the affinity constant for specific binding, $K \equiv K_p^{P,S} \mathrm{x\ pM}$ and the effective affinity ratio, $r \equiv c_{RNA}^{NS} r_p^P \cdot r_p^{F,P} \approx c_{RNA}^{NS} K_p^{P,NS}/K_p^{P,S} \mathrm{x\ } 10^{-3} \mathrm{\ pM}^{-1}$ which provides a measure of the intensity ratio due to non-specific and specific hybridization at $c_{RNA}^S = 1$ pM. Typically, the mean affinity of non-specific hybridization is two to three orders of magnitude smaller than the affinity of the probes for specific transcripts (see the data given in Fig.1). On the other hand, the binding constant for specific association of the PM exceed that of the MM by a factor between about two and twenty. The PM intensity of all considered examples is therefore distinctly higher than that of the respective MM probe at high specific transcript concentrations.

The relation between the PM and MM intensities is however more heterogeneous in the limit of dominating non-specific hybridization (i.e. at small spiked-in concentrations). This result indicates that the affinity of the PM probes for non-specific transcripts is either higher, equal or even smaller compared with that of the respective MM. Note also that the binding affinities vary by nearly three orders of magnitude between the different probes especially in the limit of small specific transcript concentrations. In addition, the PM curves are shifted by different degrees relatively to the MM curves. This result indicates a puzzling relation between the affinities of the PM and MM probes due to the mismatched base pair in the middle of the probe sequence. The binding constant of each individual probe is directly related to the strengths of the base pairings in the respective DNA probe/RNA transcript duplex and thus it depends on the sequence of the probe. The consideration of the sequence given in Fig. 1 provides however no simple explanation of the observed intensity courses.

**Figs. 1 and 2**

In summary, the dependence of the signal intensity of PM and MM probes on the concentration of target RNA can be well approximated by binding isotherms of the Langmuir type which are characterized by the binding constants of specific and non-specific hybridization for each individual probe. The whole ensemble of about 250.000 PM and MM probes on the HG U133 chip consequently requires the knowledge of nearly $10^6$ affinity constants to predict their intensity as a function of the concentration of target RNA. The determination of this rather high number of constants by model fits of the binding isotherms appears hardly to realize because one needs spiked-in data for each probe.

**Mean binding isotherms**

The mean PM and MM probe intensities, which are log-averaged over the ensemble of spiked-in genes are shown in Fig. 2 as a function of the spiked-in concentration (symbols). These data illustrate the basic effect of the transcript concentration on the probe intensities. The PM and MM probes on the average possess similar intensities at small specific transcript concentrations, i.e. at dominating non-specific hybridization. In the intermediate concentration range both, the PM and MM intensity almost linearly increases with increasing specific transcript concentration. The vertical shift between the PM and MM data reflects the mean affinity difference for specific hybridization between the PM and MM probes. Upon further increasing concentration the experimental PM intensity data progressively deviate in negative direction from the linear relationship owing to the onset of saturation. Note that the mean MM intensity values are considerably less affected by this effect.

The mean intensities are well described by Eq. 3 with $\Sigma$=sp-in (see lines in Fig. 2). It turns out that the mean binding constant of the PM probes for target RNA exceeds that of the MM on the average by a factor of $K_0^{PM,S}/K_0^{MM,S} \approx 6$-7. The effective affinity of the PM for specific binding is by two-three orders of magnitude stronger than that for non-specific binding ($r_0^{PM} = 0.0035$). On the other hand the mean affinity of PM and MM probes for non-specific binding is equal in magnitude.

**The sensitivity profiles along the probe sequence**

The nearly 250.000 PM and MM probe sensitivities per chip were analysed in terms of the position dependent single base (SB) model (Eq. 8) in correspondence with recent studies [20,21]. This approach quantifies the individual, sequence-specific intensity of each probe as the deviation from the respective set-average. Accordingly, the formation of probe-target hybrid duplexes is described by four SB sensitivity parameters for each position of the 25meric probes. The set of positional dependent sensitivity coefficients providing the optimal fit of the sensitivity values of all PM and MM probes of the HG U133 chip are shown in Fig. 3. The PM sensitivity profiles of base C and A change in a parabola-like fashion along the probe sequence being maximum and minimum in the centre at k=13, respectively. The substitution of an A by a C at position k=13 is expected to enhance the probe sensitivity by the factor $\sim 10^{0.4} = 2.5$. Note that the intensity of a poly C probe is about $10^5$ times higher than that of a 25meric poly A.

**Fig. 3**

Contrarily, the sensitivity terms for G and T monotonously change along the sequence. Differences between the base specific sensitivities almost completely vanish at the free 5' end of the probe at k=25 whereas the sensitivity of G is considerably larger than that for T at the 3' end, which is attached to the glass slide. Note that the bases G and T provide only tiny contributions to the positional dependent base sensitivity in the centre of the sequence at k=13. The base-specific sensitivity profiles are nearly equal for PM and MM probes except the small "dents" in the middle of the MM sequence for A and C and their slightly larger absolute values.

The position-dependence of the sensitivity terms can be rationalized by a gradient of the base specific contribution to the free energy of base-pair interactions along the sequence. For example, the higher flexibility of the oligonucleotide chain near its free end is expected to reduce the base specificity owing to entropic effects. On the other hand, it should be taken into account that the positional dependent SB contributions are mean parameters, which are averaged over all individual DNA/RNA duplexes of one spot. Each microscopic state contributes to the SB sensitivity with a weight according to the probability of occurrence of the respective base pairing in the dimers. Consequently also "zippering effects", e.g., target/probe duplexes which look like a partly opened double-ended zipper [25], and/or shorter probe lengths with less than 25 bases due to imperfect synthesis [26,27] potentially cause a gradient of sensitivities along the sequence because the probability of paired bases is expected to decrease in an asymmetrical fashion in direction towards the 3' and 5' ends of the oligonucleotide probe.

**The effect of specific and non-specific hybridization on the sensitivity profile**

### Figs. 4 and 5

The LS experiment enables us to study the effect of the probe sequence on the sensitivity as a function of transcript concentration. Figure 4 shows the log intensities (panel above) and the respective sensitivities (panel below) of the PM and MM probes of the spiked-in genes at selected concentrations as a function of the set averaged intensity, $<\log I_p^P>_{set}$. With increasing spiked-in concentration the data clouds shift in direction of higher abscissa values. The progressive shift between the PM and MM values reflects the higher affinity of the PM probes for specific binding (see also Fig. 2). Note that the intensity values increase with increasing concentration of specific transcripts whereas the respective sensitivity is virtually independent of the amount of spiked-in transcripts.

The sensitivity data are fitted by means of the SB model for each concentration. Note that the spiked-in data set of 3(number of genes per concentration)*14(number of concentrations)*11(number of probes per set) = 462 probes enables the determination of the 100 positional dependent sensitivity coefficients, $\sigma_k^P(B)$ (k=1,…25; B=A,T,G,C) for P=PM and MM probes. The respective sensitivity profiles (Fig. 5) are distinctly more noisy owing this relatively small number of used intensity data than the profiles which have been obtained by the fit of all, nearly 250.000 probes pairs per chip (compare with Fig. 3).

### Fig. 6

The comparison of the sensitivity profiles reveals that their distribution width about the abscissa (see, e.g., the difference $\sigma_{13}^P(C) - \sigma_{13}^P(A)$ in the middle of the sequence) progressively decreases with increasing transcript concentration. We calculated the base and position averaged absolute value of the sensitivity terms, $\sigma^P$ (Eq. 11), to quantify the observed tendency (see Fig. 6, thick lines, left ordinate). In addition we determined the mean absolute sensitivity value, $<|Y^P|>_{c=const}$, for each concentration (see symbols in Fig. 6), which characterizes the variability of the probe intensities with respect to their set

average. Note that $\sigma^P$ specifies the variability of the probe sensitivity due to the heterogeneity of the sequence in contrast to $<|Y^P|>_{c=const}$, which includes also sequence-independent contributions. The parallel change of $\sigma^P$ and $<|Y^P|>_{c=const}$ indicates that the increase of variability with decreasing transcript concentration is related to the sequence and thus to changes of the effective affinity of target/probe duplex formation.

The sensitivity is directly related to the variability of the probe intensity in a logarithmic scale, $\delta \log I^P$ (see Eq. 6). Differentiation of Eq. 1 at $c_{RNA}^S$=const and F=const and assuming $S^P$=1 for sake of simplicity provides $\delta \log I^P$ as a function of the variability of the effective binding constant of specific and non-specific transcripts in a logarithmic scale, $\delta \ln K_p^{P,h} \approx \delta K_p^{P,h}/K_0^{P,h}$, i.e.,

$$|Y^P| \propto \delta \log I^P \approx \frac{\sqrt{a+\left(\delta I^P\big|_{c=const}\right)^2}}{\ln 10 \cdot \langle I^P \rangle} \quad \text{with}$$

$$\delta I^P\big|_{c=const} = F \cdot K_0^{P,S} \cdot \left(c_{RNA}^S + c_{RNA}^{NS} \cdot r_0^P \cdot r_\Delta^P\right) \cdot \delta \ln K_p^{P,S} \quad , \tag{16}$$

$$F = F_0 \cdot \langle N_p^{F,S} \rangle \quad \text{and} \quad r_\Delta^P = \frac{\delta \ln K_p^{P,NS}}{\delta \ln K_p^{P,S}}$$

Note that the form of Eq. 16 is compatible with the error model used for the weighting factor of the least squares fits with b=0 and c=$\delta I^2$ (see above). The additive term "a" refers to fluctuations of the optical background, of the concentration and composition of the RNA as well as of the chip-specific factor. The thinner lines in Fig. 6 are calculated by means of Eq. 16. Their courses reasonably agree with the experimental data.

The logarithmic scale of the binding constant used in Eq. 16 is directly related to the free energy of binding (see Eq. 4). Hence, Eq. 16 is justified if one assumes variations of the probe intensity, which linearly scale with the free energy of duplex formation. The ratio $r_\Delta^P$ specifies the variability of the binding affinity of non-specific transcripts relatively to that of specific ones. Note that $K_p^{P,NS}$ represents an effective binding constant referring to a cocktail of RNA fragments which bind non-specifically to the probes in contrast to $K_p^{P,S}$, which is the binding affinity of the single target sequence. It is therefore reasonable to assume $\delta \ln K_p^{P,NS} > \delta \ln K_p^{P,S}$, i.e. a higher variability of the affinity for non-specific transcripts due to their more heterogeneous base composition. Note that the error model considers only "stochastic" effects in replicated measurements whereas the variability data shown in Fig. 6 (and Eq. 16 ) in addition include systematic contributions due to variations of the affinity between probes of different sequences.

We conclude that the inflation of the variability of the sensitivity (and the probe intensity) at small concentrations of specific transcripts (and at small set-averaged intensities) is partially caused by a higher variability of the binding affinity of non-specific transcripts compared with that of specific ones. The higher variability of the sensitivity of the MM in the asymptotic range at higher abscissa values reflects the higher relative contribution of variations of the binding constant, $\delta K_p^{MM,S}/K_0^{MM,S} > \delta K_p^{PM,S}/K_0^{PM,S}$.

**The sensitivity of matched and mismatched base pairings**

**Figs. 7, 8 and 9**

To compare the position-dependent sensitivity profiles at different transcript concentrations we separately plot their normalized values, $\sigma_k^P(B)_{rel} = \sigma_k^P(B)/\sigma^P$, in Figs. 7 and 8 for P=PM and MM, respectively. The PM-profiles of each base are virtually invariant with changing concentration of specific transcripts. Hence, non-specific and specific hybridization can be well described by almost the same set of relative sensitivity terms in a first-order approximation. This result is confirmed by the observation that the sensitivity profiles of the reduced ensemble of spiked-in probes scatter about the respective $\sigma_k^P(B)_{rel}$ profile obtained from the full ensemble of all probes of the chip (see circles in Fig. 7).

For the MM profiles the results dramatically change at position k=13 of the probe sequence, which refers to the mismatched self-complementary pairing with the target RNA sequence. The absolute value of the SB sensitivity contributions of the middle bases A and C progressively decreases with increasing concentration of specific transcripts (see Fig. 9). Their specific contribution to the probe sensitivity almost completely vanishes at spiked-in concentrations greater than 128 pM. Note that the bases T and G provide only tiny values of the SB sensitivity terms at position k=13 at all concentrations. Hence, the sensitivity of the MM probes is virtually invariant with respect to the mismatched base in the middle of the sequence if specific transcripts dominate hybridization. In other words, the middle base provides essentially no base-specific contribution to the stability of the duplex. On the other hand, the nearly linear relation between the MM probe intensity and the spiked-in concentration strongly indicates that the target RNA "specifically" binds to the MM probes (see Fig. 2). This result lets us conclude that specific binding to the MM probes is mainly driven by the remaining bases at positions k=1…12 and 14…25, which enable duplex formation via Watson-Crick base pairings.

# Discussion

We studied the probe intensities of Affymetrix GeneChips as a function of the concentration of specific transcripts, the sequence of which completely matches the respective PM probe sequence by complementary bases. Specific hybridization is typically overlaid by non-specific hybridization. Non-specific RNA transcripts only partially match the probe sequence by WC pairings.

The concentration dependence of the signal intensity of each probe can be well described by a simple two-state Langmuir hybridization isotherm, which considers the binding equilibria between free and bound species of specific and non-specific transcripts (see Eq. 1). In our approach all free RNA fragments compete for duplex formation with the binding sites provided by the oligonucleotide probes. It turns out that the binding of non-specific transcripts to MM probes is on the average characterized by a similar mean binding constant when compared with that of the PM probes, $K_0^{PM,NS} \approx K_0^{MM,NS}$. Contrarily, the affinity of the MM for specific transcripts is on the average nearly one order of magnitude smaller that that of the PM, $K_0^{PM,S} > K_0^{MM,S}$. The relations between the binding affinities can be summarized as PM(specific)>MM(specific)>>PM(non-specific)≈MM(non-specific).

The deviation of the intensity of an individual probe from its mean value over an appropriately chosen ensemble of probes in the logarithmic scale defines its sensitivity. It can be described as the sum of positional and base dependent terms, $\sigma_k^P(B)$ (see Eq. 8), in accordance with previous models [20,21,28,29]. Our results show that the PM-sensitivity profile is virtually independent of the concentration of target RNA, $c_{RNA}^S$. Hence, non-specific hybridization dominating at small $c_{RNA}^S$-values and specific hybridization dominating at high $c_{RNA}^S$-values give rise to virtual identical profiles of the PM-sensitivity terms. This result surprises if one considers the large difference between the mean binding constant for specific and non-specific hybridization of more than two orders of magnitude. Also the MM profiles closely resemble that of the PM for all sequence positions except the middle base. It turns out that the middle position of the MM only weakly contributes to its sensitivity.

The sensitivity terms, $\sigma_k^{P,h}(B)$, decompose into contributions due to the binding affinity, $\Delta\varepsilon_k^{P,h}(B)$, and fluorescence emission, $\Delta\varphi_k^{P,h}(B)$, according to Eq. 14. The fluorescence provides only a relatively small contribution of $|\Delta\varphi_k^{P,h}(B)| \leq 0.04$ to the SB sensitivity terms at least in the middle of the sequence ($|\sigma_{13}^{P,h}(B)| < 0.15$ for B=C,A; see Fig. 3). In other words, the observed probe sensitivity mainly reflects the sequence specific affinity for duplex formation, i.e., the propensity of the probe to bind RNA fragments from the hybridization solution. Hence, the sensitivity terms can be interpreted to a good approximation by the respective incremental contributions to the interaction free energy, i.e., $\Delta\varepsilon_{13}^{P,h}(B) \approx \sigma_{13}^{P,h}(B)$. In the following we discuss the obtained results using this approximation.

**Base pair interactions in specific duplexes**

The PM probe and the RNA target match each other via complementary Watson-Crick (WC) base pairs. The respective positional dependent free energy terms (see Eq. 5) consequently characterize the binding strength of WC pairings in the specific duplexes, i.e., $\varepsilon_{0,k}^{PM,S} \approx \varepsilon_{0,k}^{WC}$ and $\Delta\varepsilon_k^{PM,S}(\xi_{p,k}^{PM}) \approx \Delta\varepsilon_k^{WC}(\xi_{p,k}^{PM})$. Also the MM probes bind the specific transcripts via WC pairs except the middle base at position k=13, which faces "itself" in a self complementary (SC) pair (see Fig. 10 for illustration). One can therefore expect that the positional dependent free energy terms of the PM and MM probes are nearly identical for k≠13 but different for k=13. The binding strength of the middle base of the MM consequently refers to the SC pairing, i.e. $\varepsilon_{0,13}^{MM,S} \approx \varepsilon_{0,13}^{SC}$ and $\Delta\varepsilon_{13}^{MM,S}(\xi_{p,13}^{MM}) \approx \Delta\varepsilon_{13}^{SC}(\xi_{p,13}^{MM})$.

**Fig. 10**

The fit of the SB model to the sensitivity data referring to large spiked-in concentrations provides estimates of the incremental free energy of duplex stabilization (see Eq. 13). We obtained similar values for PM and MM outside of the middle base as expected, $\sigma_k^{MM,S}(B)|_{k\neq 13} \approx \sigma_k^{PM,S}(B)|_{k\neq 13} \approx$ $\Delta\varepsilon_k^{WC}(B)|_{k \neq 13}$ (see Figs. 5, 7 and 8). The relatively small contribution of the middle base of the MM, $|\sigma_{13}^{MM,S}(B)| \approx |\Delta\varepsilon_{13}^{SC}(B)| \leq 0.05$, indicates that the SC pairings on the average have virtually lost their sensitivity. This result is compatible with $|\Delta\varepsilon_{13}^{SC}(B)| << |\Delta\varepsilon_{13}^{WC}(B)|$ for B=A,C and with $|\Delta\varepsilon_{13}^{SC}(B)| \approx |\Delta\varepsilon_{13}^{WC}(B)| \approx 0$ for B = G,T. One obtains for the binding constant of the MM (see Eqs. 4 and 5)

$$-\log K_p^{MM,S} \approx \sum_{k \neq 13}^{N_b} \varepsilon_k^{WC}(\xi_{p,k}^{PM}) + \varepsilon_{13}^{SC}(\xi_{p,13}^{MM}) = -\log K_p^{PM,S} - \varepsilon_{13}^{WC-SC}(\xi_{p,13}^{PM}) \qquad (17)$$

Equation 17 shows that the log-difference of the binding constants of the PM and MM probes roughly estimates the free energy difference between the respective WC and SC pairs in the middle of the probe sequence, $\varepsilon_{13}^{WC-SC} = -(\log K_p^{PM,S} - \log K_p^{MM,S})$.

After decomposition of the free energy term according to Eq. 5 one obtains for the mean difference $\varepsilon_{0,13}^{WC-SC} = -(\log K_0^{PM,S} - \log K_0^{MM,S})$ and $\Delta\varepsilon_{13}^{WC-SC}(B) \approx \Delta\varepsilon_{13}^{WC}(B) \approx -\sigma_{13}^{PM,S}(B)$ for the base specific increment between the free energy of a WC and SC pairing in RNA-target/DNA-probe duplexes on the microarray. The fit of the total mean intensities of all spiked-in probes (Fig. 2) provides $-\varepsilon_{0,13}^{WC-SC} \approx 0.85 \pm 0.04$ (see also Table 1). This value well agrees with the mean reduced Gibbs free energy of a WC pair in DNA/RNA oligonucleotide duplexes in solution, $-\varepsilon_{sol}^{WC} = 0.75-1.03$, which was estimated using literature data of the respective nearest neighbor free energy terms [5,30] (see footnote in Table 1). The agreement between the microarray and solution data can be rationalized if the mean free energy contribution of the SC pairs to duplex stability is on the average much weaker than the contribution of the WC pairs, $|\varepsilon_{0,13}^{SC}| << |\varepsilon_{0,13}^{WC}|$ and if the contribution of the WC pairs outside of the middle base is similar in specific and non-specific duplexes.

Taking together we found that the SC pairs, on the average, only weakly contribute to the stability of probe-target duplexes. This result gives rise to $\varepsilon_{13}^{WC-SC} \approx \varepsilon_{13}^{WC}$ or equivalently $|\varepsilon_{13}^{SC}| << |\varepsilon_{13}^{WC}|$. In other words, the free energy difference between the WC and SC pairs roughly reflects the strength of the respective WC pairing in the duplexes.

**Base pair interactions in non-specific duplexes**

By non-specific binding we imply the ensemble of lower affinity mismatched duplexes involving sequences other than the intended target. The fit of the SB model to the spiked-in data at small concentrations of specific transcripts provides positional dependent sensitivity terms, which to a good approximation agree with the respective free energy contributions to the stability of non-specific duplexes (see above). For the PM and MM probes we found very similar values, which in turn also agree with the respective PM data for the specific transcripts. The latter values have been assigned to WC pairings between the probe and the RNA fragments. This agreement confirms the expectation that the non-specific duplexes are mainly stabilized by WC pairings. Consequently also the middle base in the non-specific dimers of the MM usually forms a WC pair in contrast to the respective specific duplexes where the middle base forms a SC pair (see Fig. 10 for illustration). This assignment of base-pairings in non-specific duplexes appears plausible because the "cocktail" of non-specific RNA fragments in the hybridization solution usually contains enough sequences, which enable WC pairings with the central base of the MM and the complementary central base of the PM as well. Note that the middle base of the MM is per definition a "mismatched SC" pair only with respect to the respective target sequence that specifically hybridizes the probe but not with respect to non-specific RNA-fragments.

One obtains for the binding constant of the MM after consideration of the respective relations between the free energy parameters and of Eqs. 4 and 5

$$-\log K_p^{MM,NS} \approx \sum_{k \neq 13}^{N_b} \varepsilon_k^{WC}(\xi_{p,k}^{PM}) + \varepsilon_{13}^{WC}(\xi_{p,13}^{MM}) = -\log K_p^{PM,NS} - \varepsilon_{13}^{WC-WC}(\xi_{p,13}^{PM}) \qquad (18)$$

Equation 18 shows that the log-difference between the binding constants of the PM and MM probes estimates the Gibbs free energy difference between their complementary WC pairs in the middle of the probe sequence. After decomposition of the energetic term according to Eq. 5 one obtains the base independent mean difference $-\varepsilon_{0,13}^{WC-WC} = \log K_0^{PM,NS} - \log K_0^{MM,NS}$ and the difference of the base-dependent increment $\Delta\varepsilon_{13}^{WC-WC}(B) = -\Delta\varepsilon_{13}^{WC-WC}(B^c) \approx \sigma_{13}^{PM,NS}(B) - \sigma_{13}^{PM,NS}(B^c)$ ($B^c$ denotes the complementary base of B, see also Table 1).

With Eqs. 17 and 18 one obtains for the ratio of the binding constants of non-specific and specific hybridization of each PM/MM probe pair

$$\log r_p^{PM} = \log r_0^{PM} = \log K_0^{PM,NS} - \log K_0^{PM,S} \qquad \text{and}$$

$$\log r_p^{MM} = \log r_0^{MM} - (\Delta\varepsilon_{13}^{WC}(B) - \Delta\varepsilon_{13}^{SC}(B)) \qquad (19)$$

with $\log r_0^{MM} = \log K_0^{MM,NS} - \log K_0^{MM,S} = \log r_0^{PM} - \varepsilon_{0,13}^{WC-SC}$. Note that the ratio of the PM, $r_p^{PM}$, is independent of the middle base because it forms WC pairings in specific and non-specific duplexes as well. Consequently the base-specific effect cancels out. Contrarily, the ratio for the MM depends on the middle base. Here the central WC pair in the non-specific duplexes is replaced by a SC pairing in the specific dimers.

Rearrangement of Eq. 19 provides

$$\log K_p^{P,NS} \approx \log K_p^{P,S} + \log r_0^{PM} + \begin{cases} 0 & for \quad P = PM \\ -\varepsilon_{13}^{WC-SC}(B) & for \quad P = MM \end{cases} \qquad (20)$$

The second term is either a constant (P=PM) or it depends only on the middle base (MM). It consequently does not affect the obtained sensitivity profiles at all positions (PM) or at all positions except the middle base (MM) because the symmetry condition (Eq. 9) cancels out constant contributions. This result explains the very similar base and positional dependent SB sensitivity profiles of non-specifically and specifically hybridized probes.

The stability of non-specific probe/target duplexes of the PM and MM is governed by WC pairings according to this interpretation. Consequently PM and MM probes with the same middle base are expected to hybridize with non-specific transcripts on the average almost equally. For randomly distributed middle bases one expects a vanishing mean difference, $\varepsilon_{0,13}^{WC-WC}|_{random}=0$. The fit of the total mean intensities of all spiked-in probes however provides $-\varepsilon_{0,13}^{WC-WC} \approx 0.05 \pm 0.04$ (Fig. 2, Table 1). This bias can be, at least partially, explained by a non-random distribution of middle bases for the probes on the chip according to

$$\log K_0^{P,NS} = \langle \log K_B^{P,NS} \rangle_{chip} = \langle \log K_B^{P,NS} \rangle_{random} - \sum_{B=A,T,G,C} f_{13}^{chip}(B) \cdot \Delta \varepsilon_{13}^{WC}(B),$$

where $f_{13}^{chip}(B)$ denotes the fraction of middle base B in all probes of the chip (see Table 1). With $\langle \log K_B^{PM,NS} \rangle_{random} = \langle \log K_B^{MM,NS} \rangle_{random}$ one obtains after some rearrangements

$$-\varepsilon_{0,13}^{WC-WC} = \log K_0^{PM,NS} - \log K_0^{MM,NS} = -\sum_{B=T,C} \Delta f_{13}^{chip}(B) \cdot \Delta \varepsilon_{13}^{WC-WC}(B) \quad . \quad (21)$$

$$\text{with} \quad \Delta f_{13}^{chip}(B) = \left( f_{13}^{chip}(B) - f_{13}^{chip}(B^c) \right) \quad \text{and} \quad \Delta \varepsilon_{13}^{WC-WC}(B) = \Delta \varepsilon_{13}^{WC}(B) - \Delta \varepsilon_{13}^{WC}(B^c)$$

With the respective $f_{13}^{chip}(B)$-data (see Table 1) one obtains $-\varepsilon_{0,13}^{WC-WC} \approx 0.04$ in agreement with the observed value. Note that in addition, also a non-random base distribution within the non-specific RNA fragments in the hybridization solution can introduce an asymmetry between the respective PM and MM intensities.

**Middle base averaged hybridization isotherms**

It is well established that the middle base systematically affects the relation between the PM and MM probe intensities [6], which, in addition, changes as a function of specific transcript concentration [31]. This characteristic behavior can be understood in the light of the molecular hybridization theory presented in the preceding sections.

Particularly, the SB model predicts that the relation between the intensities of the PM and MM probes depends in a characteristic fashion on the middle base (see Eqs. 17 and 18). To further check this result we calculated mean values over all spiked-in probes with a common middle base B as a function of transcript concentration. Equation 1 predicts for the middle base averaged log intensity the isotherm

$$\log I_B^P \equiv \langle \log I^P \rangle_B \approx \log F_0 + \log N_B^{F,S} + \log K_B^{P,S} + \log \left[ c_{RNA}^S + c_{RNA}^{NS} \cdot r_B^P \cdot r_B^{F,P} \right] - \log S_B^P$$

$$\text{with} \quad \log S_B^P \equiv \langle \log S^P \rangle_B \approx \log \left( 1 + K_B^{P,S} \cdot \left[ c_{RNA}^S + c_{RNA}^{NS} \cdot r_B^P \right] \right) \quad . \quad (22)$$

The effective binding constants in Eq. 22 are averages over all probes with the respective middle base B,

$$\log K_B^{P,h} \equiv \langle \log K_p^{P,h} \rangle_B \approx \log K_0^{P,h} - \Delta \varepsilon_{13}^{P,h}(B) \quad \text{with} \quad h = S, NS \quad,$$

$$\log r_B^P \equiv \langle \log r_p^P \rangle_B \approx \log r_0^P - \left( \Delta \varepsilon_{13}^{P,NS}(B) - \Delta \varepsilon_{13}^{P,S}(B) \right) \approx \begin{cases} \log r_0^{PM} & \text{for} \quad P = PM \\ \log r_0^{MM} - \Delta \varepsilon_{13}^{WC}(B) & \text{for} \quad P = MM \end{cases} \quad . \quad (23)$$

$$\text{with} \quad B \equiv \xi_{p,13}^P = A, T, G, C$$

Here we assume that the averaging ($\langle ... \rangle_B$) cancels out all positional dependent terms with $k \neq 13$, e.g. $\left\langle \sum_{k=1}^{N_b} \varepsilon_k^{P,h}(\xi_{p,k}^P) \right\rangle_B \approx \varepsilon_{13}^{P,h}(B)$. The mean effect of labelling is characterized by the equations

$$\log N_B^{F,S} \equiv \langle \log N_p^{F,S} \rangle_B \approx \log N_0^F + \Delta \varphi_{13}^{PM,S}(B) \quad \text{and}$$

$$\log r_B^{P,F} \equiv \langle \log r_p^{P,F} \rangle_B \approx \left( \Delta \varphi_{13}^{P,NS}(B) - \Delta \varphi_{13}^{P,S}(B) \right) \approx \begin{cases} 0 & \text{for} \quad P = PM \\ 2\Delta \varphi_{13}^{WC}(B) & \text{for} \quad P = MM \end{cases} \quad (24)$$

**Figs. 10, 11**

Figure 11 compares the measured with the calculated middle-base averaged mean intensity values of the PM and MM probes as a function of the concentration of specific transcripts. The theoretical curves are calculated according to Eq. 22 using the mean affinity constants, $K_0^{P,h}$ and $r_0^P$, which were previously determined for the total average of the probe intensities (see Fig. 2 and Table 1). The middle-base specific model parameters, $\Delta\varepsilon_{13}^{P,h}(B)$ and $\sigma_{13}^{P,h}(B)$ are taken from the fits of the SB model (see Figs. 3, 5. 7 and 8, and Table 1) in accordance with the results presented above (see legend of Fig.11). Hence, the curves are "synthesized" using the parameter estimates from the independent approaches of the SB model and the mean intensity fits and thus they represent rather a prediction than a fit. The agreement between calculated and measured isotherms confirms the consistency of the chosen formalism and illustrates the behaviour of PM and MM intensities as a function of concentration.

The middle base averaged mean PM intensity exceeds the respective MM intensity over the whole concentration range of specific transcripts for pyrimidine middle bases C and T of the PM probes. Contrarily, for purine middle bases B=G,A the PM and MM intensity courses intersect each other with $\log I^{MM}_B > \log I^{PM}_B$ in the limit of non-specific hybridization and with the reverse relation, $\log I^{MM}_B < \log I^{PM}_B$, at higher concentrations of specific transcripts.

The middle base specific log-intensity differences, $\log I_B^{PM-MM} \equiv \log I_B^{PM} - \log I_{Bc}^{MM}$, changes from a characteristic duplet-like pattern into a pattern of triplet-like symmetry at small and high concentrations of specific transcripts, respectively (see Fig.12). Note that the model curves excellently reproduce the features of the measured middle-base averaged intensities (compare lines and symbols in Fig.12). The duplet-like symmetry at dominating non-specific hybridization reflects a pyrimidine-purine asymmetry of the interaction strength in WC pairings of DNA/RNA hetero duplexes. Particularly, the WC pairing in the middle of the sequence reverses between the PM and MM of one probe pair, i.e., B-b$^c$ for PM transforms into B$^c$-b for the respective MM probe (e.g., G-c* → C-g, see Fig. 10 for illustration). The reversal of the base pairing is accompanied by the reversal of sign of the respective free energy difference if one compares pairs with complementary B and B$^c$ in the middle of the PM sequence, i.e., $\Delta\varepsilon_{13}^{WC-WC}(B^c) \approx -\Delta\varepsilon_{13}^{WC-WC}(B)$. This symmetrical relation splits the respective affinities into two symmetric branches relative to the overall mean, namely for the purines G and A on the one-hand side and for the pyrimidines C and T on the other hand-side.

Our analysis shows that the mismatched SC pairs on the average only weakly contribute to the affinity between the MM probe and the respective RNA target. The different base pairings, namely the WC pair (B-b$^c$) for the PM and the SC pair (<u>B</u>$^c$-b$^c$) for the respective MM (e.g., G-c* → <u>C</u>-c*) give rise to $\Delta\varepsilon_{13}^{WC}(B) - \Delta\varepsilon_{13}^{SC}(\underline{B}^c) \approx \Delta\varepsilon_{13}^{WC}(B)$. The triplet-like symmetry of the log-intensity difference at dominating specific hybridization consequently reflects the interaction strengths in the central WC pairings of specific duplexes which roughly divides into three states according to C > G ≈ T > A.

**Background correction: the PM-MM difference**

The MM probes were designed with the intention of measuring the amount of non-specific hybridization, which contributes to the PM intensities. In particular, the almost identical sequence of the PM and MM probes of one pair is expected to bind non-specific transcripts with essentially identical affinity. The subtraction of the MM from the PM intensity is therefore expected to remove this "chemical background". Making use of Eqs. 22-24 and 14 we obtain for the PM-MM difference of the middle base averaged intensities

$$I_B^\Delta = I_B^{PM} - I_{Bc}^{MM} \approx F_0 \cdot N_B^F \cdot \frac{K_B^{PM,S}}{S_B^{PM}} \cdot (1 - E_B^S \cdot R_B^S) \cdot \{c_{RNA}^S + c_{RNA}^{NS} \cdot r_0^{PM} \cdot R_B\}$$

*with*

$$E_B^h = \frac{N_{Bc}^F}{N_B^F} \cdot \frac{K_{Bc}^{MM,h}}{K_B^{PM,h}} \approx \exp\left(\ln 10 \cdot (\varepsilon_{0,13}^{PM-MM,h} - 2 \cdot \sigma_{13}^{PM-MM,h}(B))\right) \quad ; \qquad (25)$$

$$R_B = \frac{(1 - E_B^{NS} \cdot R_B^S)}{(1 - E_B^S \cdot R_B^S)} \quad and \quad R_B^S = \frac{S_B^{PM}}{S_B^{MM}}$$

Accordingly, the PM-MM intensity difference is linearly related to the fraction of specific transcripts and thus to the expression degree in an analogous fashion as the intensity of the single PM probes. The proportionality constant of the PM-MM difference is however reduced by the middle-base specific factor $(1 - E_B^S) = 0.70$ (B=A), 0.85 (T), 0.85 (G) and 0.90 (C) (using the data listed in Table 1) compared with the respective proportionality constant of the PM intensity.

Note also that subtracting the MM intensity from the PM signal only partly removes the "chemical background". Its relative contribution is reduced by the factor $R_B = 0.55$ (T) and 0.50 (C), and additionally reverses sign, $R_B = -0.25$ (B=A) and -0.20 (G), compared with the non-specific contribution to the PM intensity. This result shows that the complementary middle letter of the PM and MM probes of one pair causes a base-specific bias of the affinity for non-specific hybridization, which introduces a systematic source of variability between the PM and MM signals. The negative sign of the $R_B$-values for B=A and G reflects the middle-base specific propensity for bright MM with purine middle bases in the limit of non-specific hybridization (i.e., $I_B^\Delta < 0$ for B=A, G). The question whether this background term significantly affects gene expression measures obtained from additive intensity models [32,33] and suited correction algorithms will be separately addressed.

**The "mysterious" MM**

The pairwise design of PM/MM probes on GeneChip microarrays is based on three basic assumptions derived from conventional hybridization theory [32], namely, (i) non-specific binding is identical for PM and MM probes. (ii) The mismatch reduces the affinity of specific binding to the MM. (iii) The fluorescence response per bound transcript is identical for PM and MM and for specific and non-specific hybridization as well. These assumptions seem to predict higher PM intensities compared with that of the MM for all probe pairs in contradiction to previous observations [6]. The "riddle of bright MM" for probe pairs with $I^{PM} < I^{MM}$ can be solved within the framework of conventional hybridization theory if one decomposes specific and non-specific hybridization and analyzes the probe-target interactions on the level of base pairings and, in particular, as a function of the middle base. The explicit consideration of the strength of the central base-pairings in probe/transcript duplexes refines the picture and elucidates the origin of bright MM in terms of the pyrimidine/purine asymmetry of base pair interaction strengths. As a consequence, the first assumption modifies into "(i) non-specific binding is on the average identical for PM and MM with a preference of probe pairs with a purine base in the middle of the PM sequence for bright MM and vice versa for pyrimidines". In conclusion, the "riddle of bright MM" is obviously due to some confusion about "what RNA hybridizes the probes".

**The performance of oligonucleotide probes: ideal sensitivity and specificity**

The hybridization isotherms of the DNA probes provide a natural starting point for the characterization of their performance. In the following we will discuss the probe sensitivity and specificity as two important criteria, which can be derived from the isotherms to judge the quality of a probe as reporter for the concentration of specific target RNA in a complex mixture of RNA fragments.

The sensitivity characterizes the "detection strength" of a probe. Our definition of the sensitivity (Eq. 6) is motivated by practical reasons, which allow the calculation of its value for each GeneChip probe using its intensity with a minimum of assumptions and computational efforts. The respective values estimate the actual sensitivity in a relative scale under real conditions, which include specific and non-specific hybridization and the degree of saturation as well. The sensitivity depends consequently also on the composition of the sample solution. We therefore introduce an ideal sensitivity, which estimates the potential detection strength of a probe for specific targets under ideal conditions, i.e. in the absence of non-specific RNA fragments and saturation. The slope of the binding isotherms in the linear range at dominating specific hybridization provides a suited measure of this ideal value. For the middle base averaged isotherms one obtains in the logarithmic scale (see Eq. 22),

$$Se_B^{P,S} \equiv \log\left(\frac{\partial I_B^P}{F_0 \cdot \partial c^S}\right)_{S=1} = \log\left(N_B^{F,S} \cdot K_B^{P,S}\right) \quad , \quad (26)$$

with P= PM, MM, $\Delta$.

In particular, we are interested to compare the performance of the PM with that of the MM probes and with that of the PM-MM intensity difference, $I_B^\Delta$ (Eq. 25). The respective ideal sensitivity difference relatively to that of the PM becomes with Eqs. 23-25

$$\begin{aligned}Se_B^{PM-MM,S} &= Se_B^{PM,S} - Se_{Bc}^{MM,S} \approx -\varepsilon_{13}^{WC-SC}(B)\\ Se_B^{PM-\Delta,S} &= Se_B^{PM,S} - Se_B^{\Delta,S} \approx -\log\left(1-E_B^S\right)\end{aligned} \quad . \quad (27)$$

It turns out that the specific sensitivity of the PM distinctly exceeds that of the MM by $Se_B^{PM-MM,S}$ = 0.55 (for B=A), 0.80 (T), 0.85 (G), 1.10 (C). These values refer to an intensity difference between PM and MM probes of about one order of magnitude under ideal conditions. Contrarily, the sensitivity of the PM-MM intensity difference is nearly as large as that of the respective PM probe as indicated by the small difference $Se_B^{PM-\Delta,S}$ = 0.15 (B=A), 0.08 (T), 0.07 (G), 0.04 (C).

Specific and non-specific RNA fragments compete for hybridization with the same probe. The specificity of a probe characterizes its selectivity, i.e. its power to decide between specific target RNA and the chemical background of non-specific RNA fragments. We define the specificity as the log-ratio of the probe response to specific and non-specific hybridization in the absence of saturation, i.e. (see Eq. 22)

$$Sp_B^P \equiv -\log\left(\left|\frac{\partial I_B^P(x^S=1)}{I_B^P(c^S=0)\cdot \partial c^{NS}}\right|\right)_{S=1} = -\log\left(\left|r_B^P \cdot r_B^{F,P}\right|\right) \qquad (28)$$

An ideal probe with a vanishing affinity for non-specific binding consequently possesses a $Sp_B^P$-value of infinity. Equations 23-25 provide the specificity difference between the PM and MM probes and between the PM intensity and the PM-MM intensity difference

$$Sp_B^{PM-MM} = Sp_B^{PM} - Sp_B^{MM} \approx -\varepsilon_{13}^{WC-SC}(B) + 2\Delta\varphi_{13}^{WC}(B)$$
$$Sp_B^{PM-\Delta} = Sp_B^{PM} - Sp_B^{\Delta} \approx -\log|R_B| \qquad (29)$$

The specificity difference between the PM and MM probes reveals similar values as the respective sensitivity difference (compare Eqs. 27 and 29), $Sp_B^{PM-MM}$ = 0.63 (B=A), 0.72 (T), 0.91 (G), 1.02 (C). Accordingly, the MM specificity is distinctly smaller than that of the PM. Note that a specificity difference of about unity means that the affinity for specific binding to the PM exceeds that to the MM by one order of magnitude compared with the respective affinity for non-specific binding. Contrarily, the negative values of $Sp_B^{PM-\Delta}$ = -0.61 (B=A), -0.26 (T), -0.69 (G), -0.30 (C) show that the specificity of the PM-MM intensity difference clearly outperforms the specificity of the PM.

Table 2 summarizes our evaluation of the different intensity measures based on Eqs. 27 and 29. These results might lead to the conclusion that the PM-MM intensity difference represents the optimal measure for specific RNA because it combines a nearly as high sensitivity with a distinctly better specificity compared with that of the PM on one hand-side but the much better sensitivity and specificity characteristics compared with that of the MM on the other hand-side.

**The performance of the microarray experiment: accuracy and precision**

The judgement of the performance of the probes also depends on the chosen experimental conditions and, in particular, on the RNA concentration and the RNA composition in the sample solution. The usual setup of the microarray experiment aims to estimate the differential expression in terms of a relative "fold" change, i.e. of the log-ratio of the transcript concentration of the sample of interest relative to that of an appropriately chosen reference sample, $DE^{true}$ = $\log\{c_{RNA}^S(samp)/c^S(ref)\}$. Gene expression data analysis processes the respective probe intensities, $I_P^P(samp)$ and $I_P^P(ref)$, to provide an estimate of the differential expression DE (see, e.g., ref. [9] for an overview).

The systematic deviation between this apparent and the true value, $\Delta DE=(DE - DE^{true})$, estimates the accuracy of the method. The specificity and the accuracy are closely related parameters because both depend on the relative contribution of non-specific hybridization to the total intensity. In other words, a highly specific intensity measure is expected to provide also highly accurate DE values. Methods that use only the PM intensity typically underestimate the differential expression by more than 30%, i.e., $\Delta DE/DE_{true} > 0.3$ [9], partly because of incomplete background subtraction. Here one expects that, e.g., the PM-MM intensity difference provides a better alternative compared with PM-only measures of DE because of its higher specificity (see above).

The precision (or resolution) of gene expression analysis characterizes the confidence level of DE, i.e., the minimum difference between two DE-values, which is judged as significant. The precision of an expression measure is inversely related to its variability, given, e.g., in terms of the standard deviation, SD(DE). Highly sensitive probes typically ensure a high precision, i.e. $SD(DE) \propto 1/Se^{P,S}$, because the relative error decreases with increasing intensity (see Eq. 6).

Our results predict a second interesting relation between the precision and the specificity of the probes besides this trivial effect. Note that the fraction of specific transcripts typically differs in the sample and the reference experiments, i.e. $c_{RNA}^S(samp) \neq c_{RNA}^S(ref)$. This change of $c_{RNA}^S$ is accompanied by an alteration of intensity according to the hybridization isotherm (Eq. 22). The specificity can be interpreted as the variation of $I_B^P$ referring to an increment of $\Delta c^S=1$, if one neglects saturation for sake of simplicity (see Eq. 28). The $Sp_B^P$-values can considerably vary as a function of the middle base. The middle base of the probes consequently introduces a systematic source of variability to the apparent differential expression between oligomers with different middle bases, which probe the RNA of the same gene. Note that the microarray probes are usually designed without special attention to their middle base. It seems appropriate to use the standard deviation of the specificity upon varying middle base as a measure of the precision of the apparent differential expression, i.e.,

$$SD(DE) \propto SD(Sp^P) \equiv \frac{1}{4}\sqrt{\sum_{B=A,T,G,C}\left(Sp_B^P - \left\langle Sp_B^P \right\rangle_B\right)^2} \qquad . \qquad (30)$$

Equations 22-23 and 29 provide for the considered intensity measures $SD(Sp^P) \approx 0.0$ (for P=PM), 0.15 (MM) and 0.19 ($\Delta$). Hence, the PM intensity should be judged as the best choice with respect to the precision of the differential expression because its specificity is invariant to changes of the middle base (see Table 2). Contrarily, the MM intensity and the PM-MM intensity difference introduce a considerable variability, which lowers the precision of the respective DE-estimates. These findings agree with the results of recent statistical analyses, which show that expression measures based on MM or PM-MM intensities are less precise than that of PM-only estimates [7]. Hence, the good performance of the PM-MM intensity difference with respect to the sensitivity and specificity of the probes (see previous section) and the accuracy of the experiment must be relativized if one takes into account the resolution of the method. On the other hand, our results show that this effect possesses a systematic origin, which is mainly due to the change of base pair interactions in the middle of the probe sequence.

Taking together we emphasize that the performance of the microarray experiment depends on the performance of the chosen intensity measures, which in turn are related to the hybridization isotherms of the probes. The explicit consideration of sequence dependent factors in combination with the concentration dependence in more sophisticated analysis algorithms is expected to improve gene expression measures.

**Summary and conclusions**

Our microscopic theory of hybridization explains the concentration dependence and the effect of the middle base on the intensity of perfect matched (PM) and mismatched (MM) microarray probes in terms of effective binding constants, which in turn depend on the base pair interactions in DNA/RNA oligonucleotide duplexes. We found that

- Both PM and MM probes bind non-specific RNA fragments on the average with similar affinity.
- Both, the PM and MM probes respond to the concentration of specific transcripts and thus to the expression degree. The mean binding constant of the PM however exceeds that of the MM by nearly one order of magnitude. The markedly weaker binding affinity of the MM can be attributed to the self complementary pairing of the middle base, which on the average only weakly contributes to the stability of the specific duplexes.
- The pyrimidine/purine asymmetry of base pair interaction in the DNA/RNA hetero-duplexes splits the intensity difference between PM and MM probes at dominating non-specific hybridization into two branches and at dominating symmetric hybridization into three branches. The former effect reflects the reversal of the central WC base pairing for each probe pair whereas the latter effect can be rationalized in terms of the relatively weak SC base pairings of the MM.
- The free energy of duplex formation between target and probe mainly determines the observed intensities whereas the heterogeneity of fluorescence labelling provides only a second order contribution.
- The PM-MM intensity difference outperforms the PM intensity in terms of specificity because it largely removes the chemical background. On the other hand, the MM signal in the PM-MM difference lowers the precision of differential gene expression measures owing to systematic effects of the middle base on the binding affinity of the MM.

In conclusion, hybridization on microarrays is in agreement with the basic rules of DNA/RNA hybridization in solution. The presented model implies the refinement of existing algorithms of probe level analysis to correct microarray data for non-specific background intensities. In particular the results suggest the consideration of a middle-base specific correction term for the PM-MM intensity difference, which takes into account the fluctuations of the background intensity due to the reversal of the WC pairing in non-specific duplexes.

**Acknowledgments**

We thank Prof. Markus Loeffler and Prof. Peter Stadler for support and discussion of aspects of the paper. The work was supported by the Deutsche Forschungsgemeinschaft under DFG grant no. BIZ 6-1/2.

**Supporting Information**

available under http://pubs.acs.org:

The supporting material addresses, (i) the single base contribution to the fluorescence emission; (ii) the signal and sensitivity error of single Affymetrix GeneChips and, (iii) provides an overview of SB free energy parameters of RNA/DNA duplexes.

**Tables**

**Table 1:** Middle-base related free energy and fluorescence contrinbutions of specific (S) and non-specific (NS) hybridization on microarrays and of DNA/RNA duplexes in solution (sol) [a]

|   | probe level | base pair level | PM middle base A | T | G | C | mean [c] |
|---|---|---|---|---|---|---|---|
| **NS** | PM [b] | WC | A-u* | T-a | G-c* | C-g | |
|  | MM [b] | WC | T-a | A-u* | C-g | G-c* | |
|  | $\Delta\varepsilon_{13}^{PM,NS}$ | $\Delta\varepsilon_{13}^{WC}$ | -0.20 | +0.05 | 0.0 | +0.25 | |
|  | $\Delta\varepsilon_{13}^{MM,NS}$ | $\Delta\varepsilon_{13}^{WC}$ | | | | | |
|  | $\Delta\varepsilon_{13}^{PM-MM,NS}$ | $\Delta\varepsilon_{13}^{WC-WC}$ | -0.25 | +0.25 | -0.25 | +0.25 | |
|  | $\varepsilon_{13}^{PM-MM,NS}$ | $\varepsilon_{13}^{WC-WC}$ | -0.15 | +0.35 | -0.15 | +0.35 | 0.05±0.04 |
|  | $\Delta\varphi_{13}^{PM,NS}$ | $\Delta\varphi_{13}^{WC}$ | +0.04 | -0.04 | +0.04 | -0.04 | |
|  | $\Delta\varphi_{13}^{MM,NS}$ | $\Delta\varphi_{13}^{WC}$ | | | | | |
|  | $\Delta\varphi_{13}^{PM-MM,NS}$ | $\Delta\varphi_{13}^{WC-WC}$ | +0.08 | -0.08 | +0.08 | -0.08 | |
| **S** | PM [b] | WC | A-u* | T-a | G-c* | C-g | |
|  | MM [b] | SC | <u>T</u>-u* | <u>A</u>-a | <u>C</u>-c* | <u>G</u>-g | |
|  | $\Delta\varepsilon_{13}^{PM,S}$ | $\Delta\varepsilon_{13}^{WC}$ | | | | | |
|  | $\Delta\varepsilon_{13}^{MM,S}$ | $\Delta\varepsilon_{13}^{SC}$ | +0.05 | +0.05 | -0.05 | -0.05 | |
|  | $\Delta\varepsilon_{13}^{PM-MM,S}$ | $\Delta\varepsilon_{13}^{WC-SC}$ | -0.25 | 0.0 | 0.05 | +0.30 | |
|  | $\varepsilon_{13}^{PM-MM,S}$ | $\varepsilon_{13}^{WC-SC}\approx\varepsilon_{13}^{WC}$ | 0.55 | 0.80 | 0.85 | 1.10 | 0.85±0.04 |
|  | $\Delta\varphi_{13}^{PM,S}$ | $\Delta\varphi_{13}^{WC}$ | | | | | |
|  | $\Delta\varphi_{13}^{MM,S}$ | $\Delta\varphi_{13}^{SC}$ | -0.04 | +0.04 | -0.04 | +0.04 | |
|  | $\Delta\varphi_{13}^{PM-MM,S}$ | $\Delta\varphi_{13}^{WC-SC}$ | 0 | 0 | 0 | 0 | |
| **sol** | | WC | A-u | T-a | G-c | C-g | |
|  | [e] | $\Delta\varepsilon_{sol}^{WC-WC}$ | -0.14/-0.23 | 0.14/0.23 | -0.21/-0.18 | 0.21/0.18 | |
|  |  | $\varepsilon_{sol}^{WC}$ | 0.52/0.62 | 0.66/0.85 | 0.96/1.24 | 1.17/1.42 | 0.75/1.03 |
|  | $f_{13}^{sp-in}(B)$ [f] | | 0.22 | 0.35 | 0.23 | 0.19 | |

[a] The "probe-level" data are deduced from the combined analysis of probe intensities in terms of the SB and the intensity/binding models (see Figs. 2 and 5). The grey background indicates independent parameters used in the model. The "base pair-level" data provide an interpretation of the "probe-level" data in terms of Watson-Crick (WC) and self complementary (SC) base pairs (see text). All free energy terms are scaled with $\sim\{-(RT\cdot\ln10)^{-1}\}$. The resolution of the chosen energy parameters was arbitrarily set to ±0.05. See also Table S in the supplementary material for definitions and relations between the parameters.

[b] Pairings of the middle base in duplexes of oligo probes with specific and non-specific RNA transcripts

[c] "mean" values are averages over the base-specific values: $A_{mean}\equiv\langle A\rangle=\frac{1}{4}\sum_{B=A,T,G,C}A(B)$, i.e., $<\varepsilon_{13}^{WC-WC}> = \varepsilon_{0,13}^{WC-WC}$ and $<\varepsilon_{13}^{WC-SC}>\approx\varepsilon_{0,13}^{WC}$ for the free energy contributions in the limiting cases of non-specific and specific hybridization, respectively.

[e] Base-specific reduced free energy contribution of DNA/RNA duplex stability in solution. The mean value for each base, $\varepsilon_{sol}^{WC}(B)=(-8RT\cdot\ln10)^{-1}\sum_{X=A,T,G,C}(G(B,X)+G(X,B))$ with T=210 K, was calculated using the respective nearest-neighbor terms taken from references [5] / [30]. The difference is $\Delta\varepsilon_{sol}^{WC-WC}=\varepsilon_{sol}^{WC}(B)-\varepsilon_{sol}^{WC}(B^c)$. The $\Delta\varepsilon_{sol}^{WC}$ and $\Delta\varepsilon_{sol}^{WC-WC}$ data should be compared with $\varepsilon_{13}^{WC}$ and $\Delta\varepsilon_{13}^{WC-WC}$, respectively (see text).

[f] fraction of PM probes with middle letter B within the ensemble of 462 spiked-in probes

**Table 2:** Summaries of the performance of GeneChip oligonucleotide probes and of the respective differential expression measures [a]

|  |  | intensity measure | | |
|---|---|:---:|:---:|:---:|
|  |  | PM | MM | PM-MM |
| **probe intensity** | **sensitivity** | + | - | + |
|  | **specificity** | - | - | + |
| **differential expression** | **accuracy** | +/- | -/+ | + |
|  | **resolution** | + | - | - |

[a]   "+" and "-" indicate good and bad performances according to Eqs. 27, 29 and 30. See text.

**Figure legends**

**Figure 1**: Binding isotherms of six selected PM/MM probe pairs (see circles and triangles, respectively) showing the log-intensity as a function of the target concentration. The data are taken from the LS data set. The curves are calculated using the binding model (Eq. 1) with $K = K_p^{P,S} \times pM$ and $r = c_{RNA}^{NS} \cdot r_p^P = c_{RNA}^{NS} \cdot K_p^{P,NS}/K_p^{P,S} \times 10^{-3}$ pM$^{-1}$ (the parameter values are given within the Figure) and $\log(F \cdot N_p^F)=4.15$. The sequence of the respective PM probe is given within each panel with an enlarged middle letter. Note the heterogeneous behavior of the different probes especially in the limit of small target concentrations.

**Figure 2**: Mean binding isotherms averaged over all different 462 PM and MM probes of the LS data set (each probe intensity was considered in triplicate). The panel below shows the log-intensity difference, $<\log I^{PM-MM}>_{sp-in} = <\log I^{PM} - \log I^{MM}>_{sp-in}$. The error bars refer to the standard deviation of the individual probe intensities. The curves are calculated by means of Eq. 3 with $K = K_0^{P,S} \times pM$ and $r = c_{RNA}^{NS} r_0^P \times 10^{-3}$ pM$^{-1}$ (see Figure for parameter values) and $\log(F)=4.15$.

**Figure 3**: Single base sensitivity profiles of PM and MM oligo probes. The profile of each base (see Figure) was obtained by the least-square fit of Eq. 8 to the sensitivities of all 248.000 probes using all 42 HG U133 chips of the LS experiment.

**Figure 4**: Intensities (panel above) and sensitivities (Eq. 6, panel below) of the spiked-in probes at three different concentrations of specific target RNA (see Figure) as a function of the set averaged probe intensity. The data clouds shift towards higher abscissa values with increasing target concentration.

**Figure 5**: Single base related sensitivity profiles of PM and MM oligo probes referring to three concentrations of specific target RNA (see figure). The profile of each base (see Figure) was obtained by the least-square fit of Eq. 8 to the sensitivities of the 462 different spiked-in probes for each concentration (each condition was realized in triplicate). Note that the width between the maxima and minima of the profiles of C and A, respectively, decreases with increasing transcript concentration owing to a decreased variability of sequence specific affinity.

**Figure 6**: Variability of the sensitivity as a function of set averaged intensity of the spiked-in probes at different concentrations of specific transcripts. The thick lines (left ordinate) are the mean absolute SB sensitivity terms, $\sigma^P$ (Eq. 11) whereas the symbols denote mean absolute sensitivity values, $<|Y^P|>$ (right ordinate). The thin lines are calculated by means of the variability model, Eq. 16 ($\delta \log(I^P)$, right

ordinate), using the mean binding isotherms (Fig. 2) and $\delta \ln K^{P,S}/r_\Delta^P = 0.52/1.70$ and $0.80/1.25$ for P=PM and MM, respectively. The increase of the variability with decreasing abscissa values can be explained by a higher variability of non-specific binding as indicated by $r_\Delta^P > 1$.

**Figure 7**: Single base related sensitivity profiles of PM oligo probes in relative units, $\sigma_k^P(B)_{rel} = \sigma_k^P(B)/\sigma^P$ referring to all 14 concentrations of specific target RNA of the LS experiment. The profiles of each base (see Figure) were obtained by the least-square fit of Eq. 8 to the sensitivities of the 462 different spiked-in probes at each concentration. The circles refer to the profiles which are obtained using all PM probes of the chip (see also Fig. 3). The profile of each base is virtually not affected by the concentration of specific transcripts.

**Figure 8**: Single base related sensitivity profiles of MM oligo probes in relative units, $\sigma_k^P(B)_{rel} = \sigma_k^P(B)/\sigma^P$ referring to all 14 concentrations of specific target RNA of the LS experiment. See legend of Fig. 7 for further details. The profile of each base is virtually not affected by the specific transcript concentration except the sensitivity term of the middle bases A and C. Their absolute value progressively decreases with increasing concentration of specific transcripts (see arrows).

**Figure 9**: Single base related sensitivity terms of the middle base of PM (thin lines) and MM (thick lines) probes as a function of specific transcript concentration. The curves for T and G are almost identical for PM and MM probes. In contrast, the middle bases C and A progressively loose their sensitivity with increasing concentration of specific transcripts (see Figure for assignments).

**Figure 10**: Base pairings in the middle of duplexes between DNA probes and RNA fragments. The non-specific (NS) duplexes are stabilized by a smaller number of WC pairings compared with the specific (S) duplexes. The middle base of the MM forms a SC pairing upon specific hybridization. Note the reversal of the WC pair in the non-specific duplexes of the PM and MM probes.

**Figure 11**: Mean binding isotherms of probe pairs with middle base B= A,T,G,C of the PM probe (see figure for assignment, squares and circles refer to PM and MM probes, respectively). The data represent averages over all spiked-in probes with the respective middle base. The curves are calculated using the binding model (Eq. 22) with the model parameters listed in Table 1, a fluorescence contribution of $\Delta^F = 0.04$ and $\log(F_0 \cdot N_B^F) = 4.15$.

**Figure 12**: Log-intensity difference of the middle base related mean binding isotherms shown in Fig.11 (see figure for assignments). The curves are calculated by means of the binding model (see legend of Fig.11 for details). The dotted curve is the mean difference averaged over all middle bases.

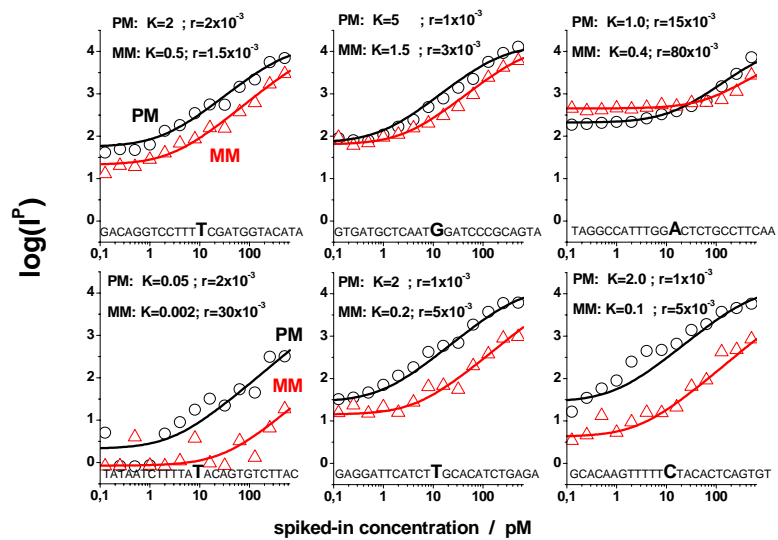

**Figure 1, Binder et al.**

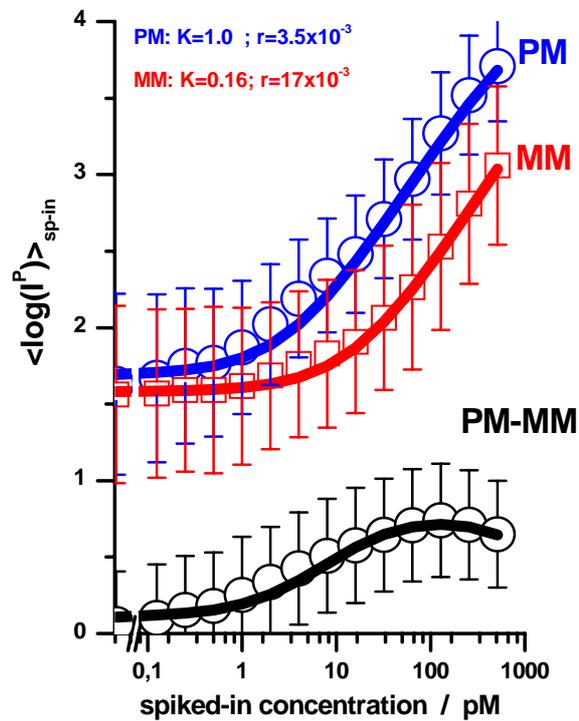

**Figure 2, Binder et al.**

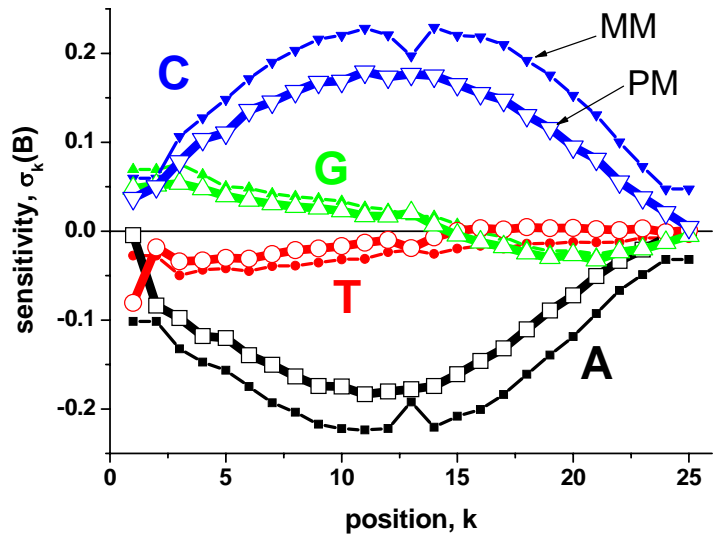

**Figure 3, Binder et al.**

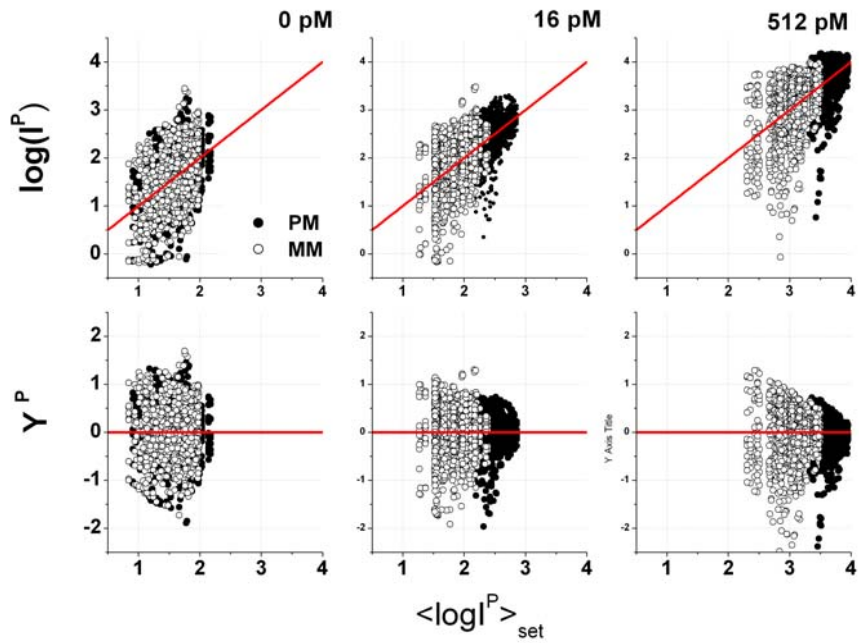

**Figure 4, Binder et al.**

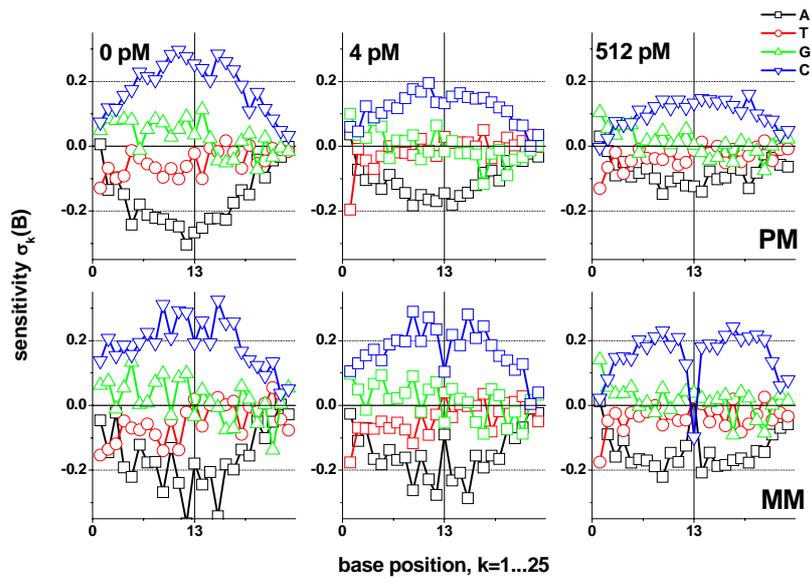

**Figure 5, Binder et al.**

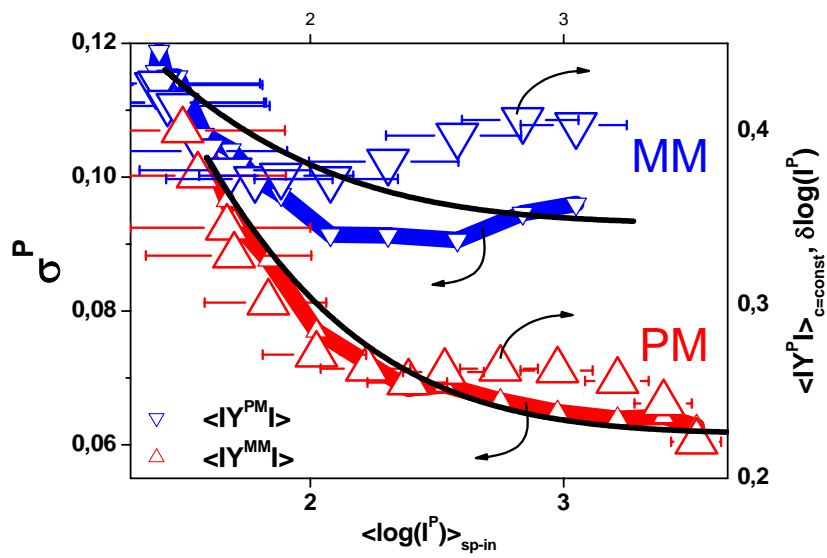

**Figure 6, Binder et al.**

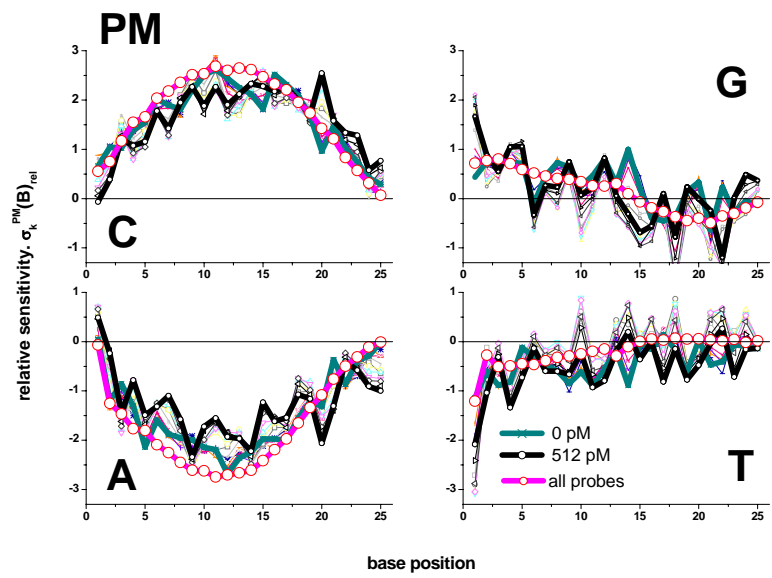

**Figure 7, Binder et al.**

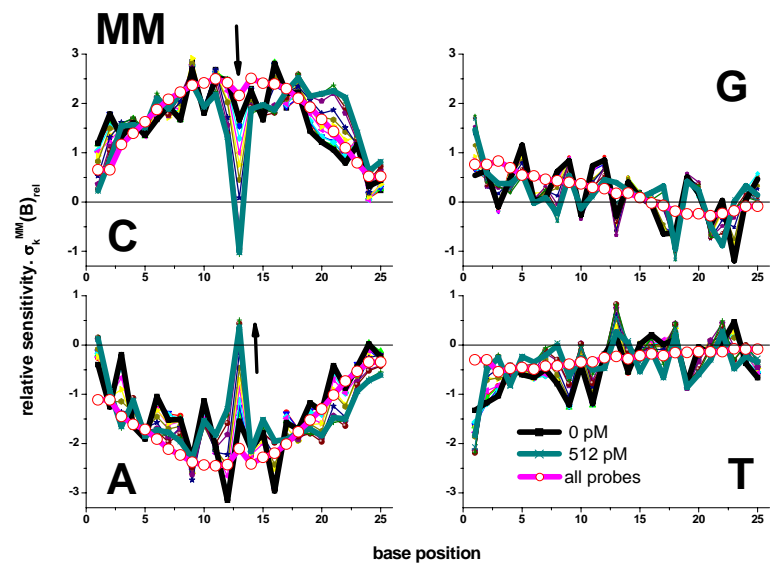

**Figure 8, Binder et al.**

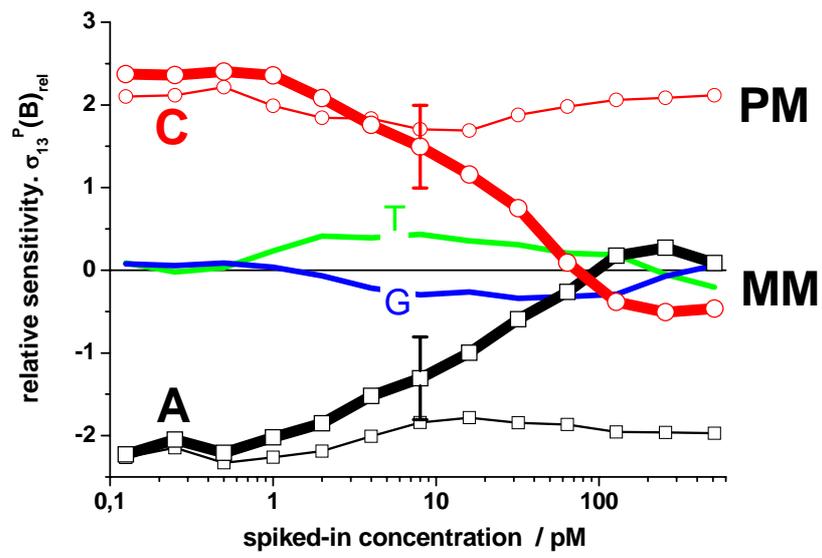

**Figure 9, Binder et al.**

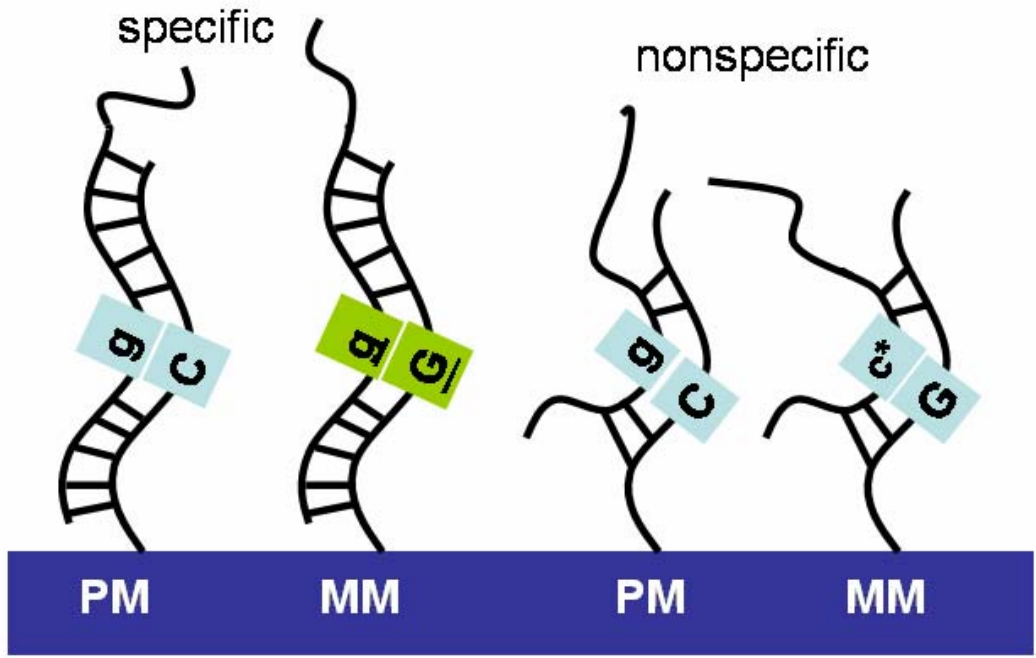

**Figure 10, Binder et al.**

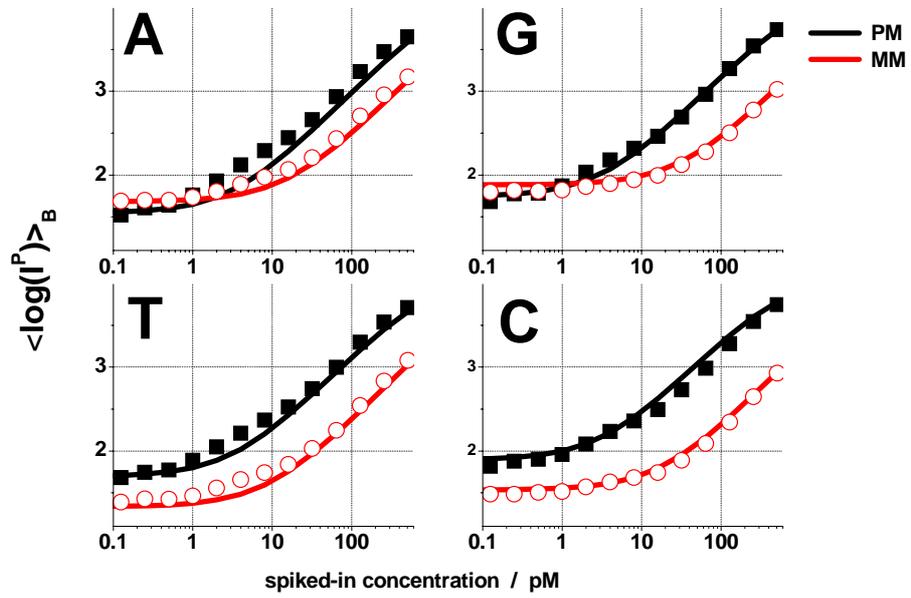

**Figure 11, Binder et al.**

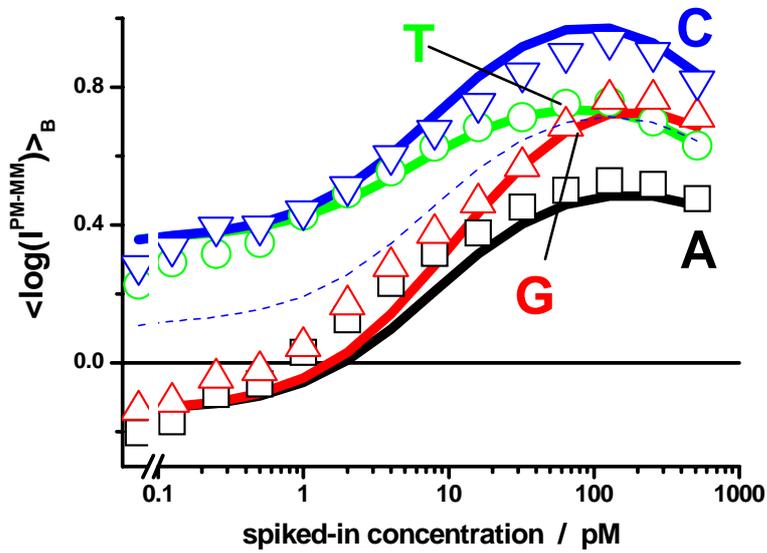

**Figure 12, Binder et al.**

# Base pair interactions and hybridization isotherms of matched and mismatched oligonucleotide probes on microarrays

Hans Binder*, Stephan Preibisch, Toralf Kirsten

Interdisciplinary Centre for Bioinformatics, University of Leipzig

* corresponding author: Interdisciplinary Centre for Bioinformatics of Leipzig University, D-4107 Leipzig, Haertelstr. 16-18, binder@izbi.uni-leipzig.de, fax: ++49-341-14951-19

# Supplementary Material

### S1: Single base contribution of fluorescence emission

The length of the RNA fragments, $N_b^{RNA}$, typically exceeds the length of the 25meric oligomer probes. Consequently also labelled bases which dangle outside of the probe/target duplex potentially contribute to the measured fluorescence intensity in addition to labels attached to the 25meric target region. Let us denote the number of bases outside of the respective 25meric duplex by $N_b^{out}$ for a RNA fragment of total length $N_b^{RNA} = N_b + N_b^{out}$. The respective number of labelled bases inside and outside of the 25mer is $N_p^{F,in}(\xi^T)$ and $N_p^{F,out}(\xi^{T,out})$, respectively, where $\xi^{T,out}$ is the subsequence of the target RNA exceeding the probe on both sides. The fluorescence intensity of a RNA fragment is related to the number of labelled c* and u*, which is given by the number of complementary G and A of the target gene according to

$$N_p^{F,S} = N_p^{F,in}(\xi^T) + N_p^{F,out}(\xi^{T,out}) = N_p^{u*}(\xi^T + \xi^{T,out}) + N_p^{c*}(\xi^T + \xi^{T,out}) = N_p^A(\xi^P + \xi^{P,out}) + N_p^G(\xi^P + \xi^{P,out})$$

if one assumes exclusively WC pairings. The contribution to the sensitivity of a selected probe owing to the number of potentially labelled bases per target, $N_p^{F,S}$, is (see Eqs. 1 and 7)

$$Y_p^{P,F} \equiv \log N_p^F - \left\langle \log N_p^F \right\rangle_{set} = \Delta_p^F \cdot \left[ \sum_{k=1}^{N_b} \sum_{B=A,G} \left( \delta(B, \xi_k^P) - f_k^{set}(B) \right) - \sum_{k=1}^{N_b} \sum_{B=T,C} \left( \delta(B, \xi_k^P) - f_k^{set}(B) \right) \right] \quad (A1)$$

with $\quad \Delta_p^F = \dfrac{\log N_p^F - \left\langle \log N_p^F \right\rangle_{set}}{\delta N_p^{F,in}} \quad$ and $\quad \delta N_p^{F,in} = N_p^{F,in} - \left\langle N_p^{F,in} \right\rangle_{set} = \sum_{k=1}^{N_b} \sum_{B=A,T,G,C} \left( \delta(B, \xi_k^P) - f_k^{set}(B) \right)$

where averaging was performed over the probe set ($\Sigma \equiv$ set).

The coefficient $\Delta_p^F$ specifies the contribution of fluorescence labelling per potentially labelled base in the considered target sequence of length $N_b$. Effectively each labelled base pair increases and each nonlabelled pair decreases the sensitivity by $\Delta_p^F$. With $\delta N_p^F = \delta N_p^{F,in} + \delta N_p^{F,out}$ ($\delta N_p^{F,i} = N_p^{F,i}$ -

$<N_p^{F,i}>_{set}$, i = in, out) and $<\delta N^F>_{set}=0$ one obtains the following approximation for $\Delta_p^F$ in the limit of small $\delta N_p^F/<N_p^F>_{set} \ll 1$, which is justified for sequence lengths $N_b^{RNA} > 20$,

$\Delta_p^F = [\log(1+\delta N_p^F/<N_p^F>_{set}) - <\log(1+\delta N_p^F/<N_p^F>_{set})>_{set}]/\delta N_p^{F,in}$

$\approx (\ln 10 \cdot \delta N_p^{F,in} \cdot <N_p^F>_{set})^{-1} (\delta N_p^F - <\delta N^F>) = (1 + \delta N_p^{F,out}/\delta N_p^{F,in})/(\ln 10 <N_p^F>_{set})$.

The binominal distribution $B(N^F, N_b^{tot}, p) = \binom{N_b^{tot}}{N^F} p^{N^F} (1-p)^{N_b^{tot}-N^F}$ specifies the probability to find $N_p^F$ potentially labelled nucleotides among a total sequence length of the target fragment of $N_b^{RNA}$ nucleotides where $p \approx 0.5$ is the probability for a uracyl or a cytosine at any position of the target sequence. After substitution of the set average by the overall mean of the number of labels per target by $<N_p^F>_{set} \approx <N_p^F>_{binom} = p \cdot N_b^{RNA}$ one gets the relative fluorescence contribution per sequence position

$$\Delta_p^F \approx \left(1 + \frac{\delta N_p^{F,out}}{\delta N_p^{F,in}}\right) \cdot \Delta_0^F \quad with \quad \Delta_0^F = \left(\ln 10 \cdot p \cdot N_b^{RNA}\right)^{-1} \approx \frac{2}{\ln 10 \cdot N_b^{RNA}} \tag{A2}$$

Its mean value,

$$\Delta^F \equiv \left\langle \Delta_p^F \right\rangle_{chip} \approx \left(1 + \sqrt{\frac{N_b^{out}}{N_b}}\right) \cdot \Delta_0^F = \left(1 + \sqrt{\frac{N_b^{RNA}}{N_b}-1}\right) \cdot \Delta_0^F \quad , \tag{A3}$$

provides the average contribution per considered base within a probe sequence of lenth $N_b$ as a function of the total length of the RNA fragment, $N_b^{RNA}$. The mean incremental contributions are approximated in Eq. A3 by the standard deviation of the binominal distribution according to $<\delta N>_{set} \approx p \cdot N^{0.5}$.

Bases in the probe sequence referring (B=A,G) and not-referring (B=T,C) to labels in the complementary target sequence add and subtract the constant contribution $\Delta^F$ to the sensitivity, respectively.

## S2: Signal and sensitivity error of single Affymetrix GeneChips

The weighting factor for the least squares fits of the positional dependent sensitivity models is given by the variance of the experimental sensitivity data, $\omega^{P\,2}_p \approx \mathrm{var}(Y^P_p)$ [1]. It can be estimated for each probe from chip replicates using standard error analysis. The SB sensitivity contributions are partly obtained from least square fits of the sensitivity data of single chips. We therefore developed a method, which estimates $\mathrm{var}(Y^P_p)$ for each individual chip using selected probe intensities.

The variance of the sensitivity can be directly related to the variance of the respective signal intensity according to Eq. 6, $\mathrm{var}(Y^P_p) \approx \mathrm{var}(\log(I^P_p)) + \mathrm{var}(<\log(I^P_p)>) \approx \mathrm{var}(\log(I^P_p))\,(1+(N_{probe}-1)^{-1}) \approx \mathrm{var}(\log(I^P_p))$ where $N_{probe} = 11 - 20$ is the number of probes per probe set. For the estimation of the chip-specific value of $\omega^{P\,2}_p$ we make use of the fact that a considerable number of PM and MM probes are present as replicates on each Affymetrix© chip. We identified repeated probes by comparison of all sequences present on the chip. For example, the human HG U133 chip contains 3463 probes in duplicate (2x), 725 in triplicate (3x), 186 fourfold (4x), 77 fivefold (5x), 37 sixfold (6x), 7 sevenfold (7x), 2 ninefold (9x) and one each for 12x, 16x and 20x. We calculated the variance, $\mathrm{var}_{exp}(\log(I^P_p))$ (P=PM, MM) and log-averaged mean intensity, $<I^P_p> = \exp(<\ln I^P_p>_{replicate})$, for each of these groups of replicates for a selected chip.

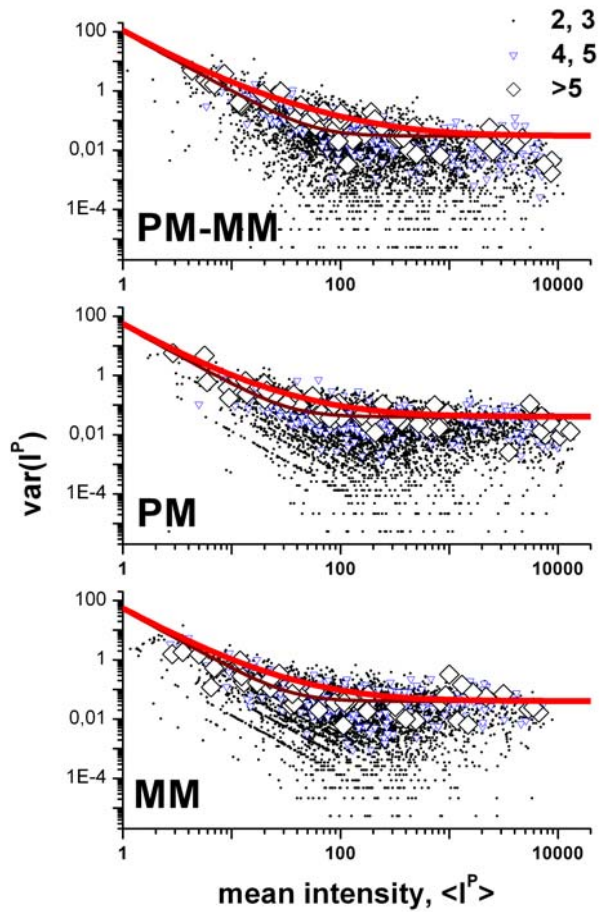

*Figure S2:* Log-log plot of the variance of the intensities of replicate PM and MM probes present on a HG U133 chip as a function of the mean intensity $<I^P>$ averaged over probes present in duplicate and triplicate (see legend in the Figure: 2,3…small points), four- and fivefold (4,5…small triangles) and more than fivefold (>5…rhombes). The lines are calculated according to the error model (Eq. A5) with a/b/c = 0.04/5/50 (thick line) and 0.04/0/50 (thin line) for P=PM and MM. The panel above shows the respective analysis of log-intensity differences, PM-MM. In this case the variance is given by $\mathrm{var}(\log I^{PM-MM}) = \mathrm{var}(\log I^{PM} - \log I^{MM})$ and the squared mean intensity by $<I^{PM+MM}>^2 = <I^{PM} \cdot I^{MM}>^{-1} = \exp[<\ln(I^{PM} + I^{MM})>]$. The thick and thin lines are calculated with a/b/c = 0.03/10/100 and 0.03/0/100, respectively.

The uncorrected signal intensity can be rewritten according to Eq. 1 (in the original article) in a simplified version as $I_p^{P*} \approx (<F_{chip} \cdot N^F \cdot c_{RNA}> + e_F) \cdot \exp[(<\ln K_p^P> + e_G)] + (<\beta_p^P> + e_B)$ where the angular brackets, <...>, denote means over replicated probes. The $e_i$ (i=F, G, B) are error terms and $\beta_p^P$ is the optical background of each probe, which is not related to hybridization. With $<I_p^P> \approx <F_{chip} \cdot N^F \cdot c_{RNA} \cdot K_p^b>$ one obtains the background-corrected intensity

$$I_p^P \approx I_p^{P*} - <\beta_p^P> \approx (<I_p^P> + e_F) \cdot \exp(e_G) + e_B \qquad (A4)$$

The constant $F_{chip}$ depends on the yield of labelling (fraction of labelled uracyls and cytosines), on the number of oligos per spot and on the efficiency of the detector and of the imaging system (see ref. [2] for details). Consequently the first error term, $e_F$, considers effects such as variations of the labelling efficiency, of the number of oligos per probe spot and of their density, of the RNA concentration and the noise of the detector and of the imaging system. The exponential term, $\varepsilon_G$, can be rationalized as the error of the free energy of duplex formation, $\Delta G_p^b \propto -\ln K_p^P$, which is related, e.g., to incorrect sequences of individual oligos in each probe spot due to imperfect synthesis and/or to non-equilibrium effects of target binding. The last error, $e_B$, considers the noise of the detector and of the imaging system in the absence of hybridization.

The variance of log transformed and background corrected signal intensity is described to a good approximation by $\text{var}_{mod}[\log(I_p^P)] \approx a + c/(<I_p^P>)^2$ with $a \approx s_G^2/(\ln 10)^2$ and $c \approx (s_F^2 + s_B^2)$ if one assumes exclusively normally distributed error terms with mean 0 and variance $s_i^2$ (i=F, G, B). This result agrees with a previously proposed error model of microarray intensity data [3].

Figure S2 compares experimental and theoretical variance data of PM and MM intensities and of their difference in a double-logarithmic scale. The model curves systematically underestimate the experimental variance data in the intermediate intensity range, $100 < I^P < 1000$ (see thin lines in Fig. S2). Considerable better agreement was achieved if one adds a term $\sim <I^P>^{-1}$ according to (see thick lines in Fig. 13)

$$\text{var}_{mod}\left(\log I_p^P\right) \approx a + \frac{b}{<I_p^P>} + \frac{c}{<I_p^P>^2} \qquad (A5)$$

The additional term can be tentatively rationalized as non-Gaussian error terms, which contribute to $e_F$. Here we use Eq. A5 without further specification as an empirical measure to estimate the weighting factor in the sum of squared residuals in the least squares fits as a function of the signal intensity.

The analysis of the intensity difference, $\text{var}(\log I^{PM} - \log I^{MM})$, as a function of $<I^{PM} \cdot I^{MM}>^{-1} = \exp[-<\ln(I^{PM} + I^{MM})>]$ provides similar plots as that for PM and MM (Fig. S2, panel above). The respective background error is however increased whereas the signal error decreases compared with the respective error data of PM and MM probes. This result is compatible with a uncorrelated background noise of PM and MM intensities. In this case one expects for the background error of the log-difference of PM and MM probes a standard deviation of $s_B^2(PM-MM) \approx s_B^2(PM) + s_B^2(MM) \approx 2s_B^2(PM)$, where the arguments PM and MM refer to the log-transformed intensities of the respective probes. On the other hand, the signal error term, "a", of $\text{var}(\log I^{PM} - \log I^{MM})$ slightly reduces when compared with that of the individual PM and MM probes. This result can be explained with correlations between the PM and MM intensities, which are discussed in the paper.

## S3: Overview of SB free energy parameters of DNA/RNA duolexes

Relations between the positional dependent SB free energy and fluorescence contributions of Watson-Crick (WC) and self complementary (SC) pairings in DNA/RNA oligonucleotide duplexes of the PM and MM microarray probes upon-specific (S) and non-specific (NS) hybridization [a]

| Du-plex | probe level | Single base contributions | | |
|---|---|---|---|---|
| | | base pair level | | |
| | Probe P= position | PM,MM $k \neq 13$ | PM $k=13$ | MM $k=13$ |
| NS | base pairing | WC: B-b$^c$ | WC: B-b$^c$ | WC: B$^c$-b |
| | $\varepsilon_{0,k}^{P,NS} \approx$ | $\varepsilon_{0,k}^{WC}$ | $\varepsilon_{0,13}^{WC}$ | $\varepsilon_{0,13}^{WC}$ |
| | $\varepsilon_{0,k}^{PM-MM,NS} \approx$ | 0 | $\varepsilon_{0,13}^{WC-WC} \approx -(\log K_0^{PM,NS} - \log K_0^{MM,NS})$ | |
| | $\Delta\varepsilon_k^{P,NS}(B) \approx$ | $\Delta\varepsilon_k^{WC}(B)$ | $\Delta\varepsilon_{13}^{WC}(B)$ | $\Delta\varepsilon_{13}^{WC}(B^c)$ |
| | $\Delta\varepsilon_k^{PM-MM,NS}(B) \approx$ | 0 | $\Delta\varepsilon_{13}^{WC-WC}(B) = -\Delta\varepsilon_{13}^{WC-WC}(B^c) \equiv \Delta\varepsilon_{13}^{WC}(B) - \Delta\varepsilon_{13}^{WC}(B^c)$ $C \approx T \approx -G \approx -A < 0$ | |
| | $\Delta\varphi_k^{P,NS}(B) \approx$ | $\Delta\varphi_k^{WC}(B)$ | $\Delta\varphi_{13}^{WC}(B)$ | $\Delta\varphi_{13}^{WC}(B^c) = -\Delta\varphi_{13}^{WC}(B)$ |
| | $\Delta\varphi_k^{PM-MM,NS}(B) \approx$ | 0 | $\Delta\varphi_{13}^{WC-WC}(B) \equiv \Delta\varphi_{13}^{WC}(B) - \Delta\varphi_{13}^{WC}(B^c) = 2\Delta\varphi_{13}^{WC}(B)$ $|\Delta\varphi_{13}^{WC-WC}(B)| \approx |\Delta^F|$ ; $G \approx A \approx -C \approx -T > 0$ | |
| | $\sigma_k^{PM-MM,NS}(B) \approx$ | 0 | $\Delta\varepsilon_{13}^{WC-WC}(B) - \Delta\varphi_{13}^{WC-WC}(B)$ | |
| S | base pairing | WC: B-b$^c$ | WC: B-b$^c$ | SC: <u>B$^c$</u>-b$^c$ |
| | $\varepsilon_{0,k}^{P,S} \approx$ | $\varepsilon_{0,k}^{WC}$ | $\varepsilon_{0,13}^{WC}$ | $\varepsilon_{0,13}^{SC}$ |
| | $\varepsilon_{0,13}^{PM-MM,S}$ | 0 | $\varepsilon_{0,13}^{WC-SC} \equiv \varepsilon_{0,13}^{WC} - \varepsilon_{0,13}^{SC} \approx -(\log K_0^{PM,S} - \log K_0^{MM,S})$ | |
| | $\Delta\varepsilon_k^{P,S}(B) \approx$ | $\Delta\varepsilon_k^{WC}(B)$ | $\Delta\varepsilon_{13}^{WC}(B)$ | $\Delta\varepsilon_{13}^{SC}(B^c)$ |
| | $\Delta\varepsilon_k^{PM-MM,S}(B) \approx$ | 0 | $\Delta\varepsilon_{13}^{WC-SC}(B) \equiv \Delta\varepsilon_{13}^{WC}(B) - \Delta\varepsilon_{13}^{SC}(B^c) \approx \Delta\varepsilon_{13}^{WC}(B)$ $C > G \approx T > A$ | |
| | $\Delta\varphi_k^{P,S}(B) \approx$ | $\Delta\varphi_k^{WC}(B)$ | $\Delta\varphi_{13}^{WC}(B)$ | $\Delta\varphi_{13}^{SC}(B^c) = -\Delta\varphi_{13}^{WC}(B^c)$ |
| | $\Delta\varphi_k^{PM-MM,S}(B) \approx$ | 0 | $\Delta\varphi_{13}^{WC-SC}(B) \equiv \Delta\varphi_{13}^{WC}(B) - \Delta\varphi_{13}^{SC}(B^c) \approx 0$ | |
| | $\sigma_k^{PM-MM,S}(B) \approx$ | 0 | $\Delta\varepsilon_{13}^{WC-SC}(B)$ | |

[a] Single base related free energy ($\varepsilon$), fluorescence ($\varphi$) and sensitivity ($\sigma$) contributions to the probe intensities. The index k indicates the position of base B=A,T,G,C along the probe sequence. k=13 refers to the middle base whereas k≠13 refers to all positions outside the middle base. The superscript "c" denotes the complementary base, e.g., for B=A one gets of B$^c$=T. Single See text.